\documentclass[10pt, conference, letterpaper]{IEEEtran}

\ifCLASSINFOpdf
\else
\fi

\usepackage{graphicx}
\usepackage{subcaption}
\usepackage{verbatim}
\usepackage{amssymb,amsmath}
\usepackage{gensymb}
\usepackage{mathtools}
\usepackage{multicol}
\usepackage{booktabs}
\usepackage{rotating} 
\usepackage{multirow}
\usepackage{epstopdf}
\usepackage{cite}
\usepackage{amsthm}
\usepackage{romannum}
\usepackage{multirow}
\usepackage{pifont}
\usepackage[affil-sl] {authblk}
\usepackage{geometry}
\usepackage{makecell}
\usepackage{authblk}
\usepackage{dsfont}
\usepackage{pifont}
\usepackage{mathtools, nccmath}
\usepackage[linesnumbered,ruled,vlined]{algorithm2e}

\newcommand{\norm}[1]{\left\lVert#1\right\rVert}
\DeclareMathSymbol{\mlq}{\mathord}{operators}{``}
\DeclareMathSymbol{\mrq}{\mathord}{operators}{`'}

\DeclareMathOperator*{\argmin}{arg\,min}

\IEEEoverridecommandlockouts
\renewcommand{\baselinestretch}{1}

\begin{document}

%
\renewcommand{\thefootnote}{\fnsymbol{footnote}}
\title{\Large Energy Minimization for Federated Asynchronous Learning on Battery-Powered Mobile Devices via Application Co-running}


\author[$\dag$,$\ast$]{\normalsize Cong Wang\thanks{* The corresponding author.}}
\author[$\S$]{\normalsize Bin Hu}
\author[$\S$]{\normalsize Hongyi Wu\vspace{-0.1in}}
\affil[$\dag$]{\normalsize George Mason University, Fairfax, VA 22030, USA}
\affil[$\S$]{\normalsize Old Dominion University, Norfolk, VA 23529, USA}
\affil[$\dag$]{cwang51@gmu.edu} \affil[$\S$]{bhu@odu.edu, h1wu@odu.edu}

\maketitle

\renewcommand{\thefootnote}{\arabic{footnote}}
\vspace{-0.1in}
\begin{abstract}
Energy is an essential, but often forgotten aspect in large-scale federated systems. As most of the research focuses on tackling computational and statistical heterogeneity from the machine learning algorithms, the impact on the mobile system still remains unclear. In this paper, we design and implement an online optimization framework by connecting asynchronous execution of federated training with application co-running to minimize energy consumption on battery-powered mobile devices. From a series of experiments, we find that co-running the training process in the background with foreground applications gives the system a deep energy discount with negligible performance slowdown. Based on these results, we first study an offline problem assuming all the future occurrences of applications are available, and propose a dynamic programming-based algorithm. Then we propose an online algorithm using the Lyapunov framework to explore the solution space via the energy-staleness trade-off. The extensive experiments demonstrate that the online optimization framework can save over 60\% energy with 3 times faster convergence speed compared to the previous schemes.
\end{abstract}



\IEEEpeerreviewmaketitle

\section{Introduction}\label{sec:intro}
Our planet is in danger. Among a variety of causes, AI plays an unequivocally  negative role to accelerate the emission of carbon dioxide and irreversible climate change. As deep learning is increasingly deployed in large-scale distributed systems, their energy footprint is growing at an unprecedented, breathtaking rate~\cite{climate}. Recently, \emph{Federated Learning} rises as a promising computing paradigm that allows participants to learn a collaborative model in privacy-preserved manner~\cite{fedavg,fednova,scaffold,fedprox}. However, by pushing neural computations to the multi-core CPUs~\cite{wang-mm}, its energy implication is far from clear on battery-powered mobile devices~\cite{carbon}: high-intensity neural computation quickly drains the battery and frequent charge/discharge also shorten the battery lifetime, and their disposal ultimately becomes an environmental liability.

The classic federated learning originates from the principles of synchronous Stochastic Gradient Descent (Sync-SGD) in cloud computing, where all the participants proceed in lock-step and their parameters are averaged at the parameter server. It is well-known that such simple migration is subject to the computational heterogeneity in mobile environments, vastly due to the segmented mobile hardware market and vendor-supplied drivers. Worst-case stragglers (slowest workers) could be orders of magnitude slower than the average execution whereas the majority of the computing power is underutilized, especially when the stragglers are experiencing heavy thermal throttling and user interference~\cite{tpds-wang}. Further, Sync-SGD does not provide the temporal flexibility of coordination and slow convergence further exacerbates the energy consumptions in the system. \emph{Asynchronous training} (ASync-SGD) is a competitive solution to tackle computational heterogeneity~\cite{hogwild} but its potential is yet to be fully explored in federated learning~\cite{xing-nips18}. ASync-SGD allows fast participants to proceed in lock-free steps while the global parameters are exchanged and kept with the most updated local ones. Without such barrier from the stragglers, more updates can be made and the wall-clock convergence time is reduced. Unfortunately, most of the research in Sync-SGD~\cite{fedavg,fednova,scaffold,fedprox} and ASync-SGD~\cite{hogwild,async-icml,async-fed,xing-nips18} lie in the confined areas of machine learning and optimization theories, but the interaction to the underlying system is not fully explored, particularly for achieving energy-efficient computation.

In this paper, we aim to optimize the energy expenditure of federated learning tasks by taking advantage of application co-running opportunities and asynchronous execution. The design stems from the pervasive ARM CPU microarchitecture, which features the big.LITTLE cores~\cite{arm} to tackle multi-tasking with energy-efficiency: the big cores of high throughput for foreground applications and the little cores of low power consumptions for system and background processes. To avoid interfering with the normal usage, the training threads can be designated as a background service and only called once a set of conditions such as networking, battery energy conditions are met. As validated in our experiments, once the training threads of high parallelization are running in the background (dispatched to the little cores by the kernel scheduler), simultaneous execution of a foreground application gives the entire system a deep energy discount (about 30-50\%) compared to running the foreground application and training separately, with negligible performance impact measured by Frame per Second (FPS). Combined with ASync-SGD, it gives the flexibility to defer gradient updates until a foreground application takes place.

A seamless integration of the system dynamics with the upper-level machine learning algorithm faces tremendous challenges. The success of asynchronous training relies on well-managed \emph{staleness} in the system, that the stale updates from the stragglers should not diverge too much from the current directions~\cite{xing-nips18,async-icml}, i.e., the staleness is bounded with low variance. Hence, the first challenge comes from the inevitable staleness in the system while waiting for better co-running opportunities, not only because of the difficulty to quantify gradient staleness, but also how to formulate them into the optimization. Second, since the patterns and future occurrences of user application are unknown, the optimization need to make real-time decisions based on the known priori. Third, how those control decisions would propagate upwards and affect the global model convergence and wall-clock training time. To tackle these challenges, we first study a basic offline scheduling problem assuming the access to all the future occurrences of the applications. We adopt a recently proposed metric called \emph{gradient gap} to measure the difference between model parameters in their norm magnitude~\cite{gap_1,gap_2}, and formulate the offline problem into a \emph{Knapsack Problem}~\cite{knapsack}, with a pseudo polynomial-time dynamic programming solution. Then we further propose an online optimization algorithm based on the Lyapunov framework~\cite{lyaponouv} to explore the two-way trade-off between energy and staleness, which in turn implies convergence speed and wall-clock training time. The framework is proved to achieve the $[\mathcal{O}(1/V), \mathcal{O}(V)]$ energy-staleness trade-off, which only requires the current information of system dynamics and queue backlogs.

The contribution of this work is many-fold. First, motivated by a series of key findings in real experiments, we leverage ASync-SGD and system-level opportunities for energy optimization of federated training tasks on edge devices. To the best of our knowledge, this is one of the few works that integrate the high-level machine learning algorithms with the low-level system dynamics on consumer's edge devices. Second, we formulate both offline and online optimization problems and design an efficient online scheduler while ensuring bounded staleness in the long term. Finally, we conduct extensive evaluations on a mobile testbed with 4 types of devices using the CIFAR10 dataset~\cite{cifar10}. The results demonstrate over 60\% energy saving compared with FedAvg~\cite{fedavg} and $3\times$ faster convergence speed.

The rest of the paper is organized as follows. Section \ref{sec:background} discusses the background and related works. Section \ref{sec:motivation} motivates this work and defines the system model. Sections \ref{sec:offline} and \ref{sec:online} describe the framework for offline and online optimization. Section \ref{sec:implementation} implements the framework on edge devices. Section \ref{sec:eval} evaluates the proposed framework and Section \ref{sec:conclusion} concludes this work with future directions. 

\section{Background and Related Works} \label{sec:background}

\subsection{Asynchronous Federated Learning}

\textbf{Sync-SGD.} The state-of-the-art Federated Learning establishes on synchronous SGD, where local workers proceed with a barrier until all the workers finish their local training~\cite{fedavg}. As pointed out in~\cite{fednova,scaffold,fedprox}, this simple migration is subject to extensive heterogeneity in mobile edge systems due to the diverse computational capability, network connectivity/bandwidth and user behaviors. \cite{fednova} develops a new aggregation rule to allow local variations such as the number of epoches and optimizers used by different participants. \cite{scaffold} analyzes the convergence instability due to stragglers and proposes a mechanism to correct those diverging gradient updates. \cite{fedprox} adds a proximal term to the objective to manage heterogeneity associated with partial information, when the straggler's updates have been dropped out. These works aim to improve the computational efficiency of Sync-SGD while preserving the statistical stability.

\textbf{ASync-SGD} is a natural way to tackle computational heterogeneity, and its original version can be traced back to \emph{HOGWILD!}~\cite{hogwild} in multicore systems. Multiple threads are allowed access to the shared memory and update the model at will. Considerable efforts have been devoted to understanding and mitigating staleness~\cite{async-icml,async-fed}. \cite{async-icml} adopts the Taylor Expansion and Hessian approximation to compensate the delay from stale gradients, while avoiding the complexities from the high-order terms. ~\cite{async-fed} introduces a regularized term to reduce the variance due to staleness. Though asynchrony introduces race conditions, it is proved to achieve optimal convergence rate at a much faster speed~\cite{hogwild}, mainly due to more number of updates are now being conducted. Some works also partially contribute the statistical efficiency to the implicit momentum introduced from the stale gradients~\cite{acc-momentum}.

Momentum plays an important role to facilitate convergence. The update is simply an exponentially weighted average that continuously adds a portion of the previous momentum vector $v_{t-1}$ to the current vector $v_t$ plus the fraction from the current gradient vector $s_t$,
\begin{equation}
\small
v_t = \beta v_{t-1} + (1-\beta) s_t, \hspace{0.05in} \theta_t = \theta_{t-1} - \eta v_t \label{momentum}
\end{equation}
then the model parameters $\theta_t$ are updated according to the learning rate $\eta$. Here, the stale gradients can be thought as the previous gradient vectors $v_{t-1}$ that can dampen oscillations along the way to the minima. Thus, it is interesting to see that the benefits and drawbacks of staleness actually co-exist, but such contradiction and its theoretical implication are still not fully understood at this stage~\cite{xing-nips18}. Most of the federated research aims at improving the Sync-SGD or ASync-SGD algorithms, but overlooks important aspects from the system, such as energy-efficiency on battery-powered edge devices. This work differs from a large body of existing works to combine ASync-SGD with system-level opportunities and reduce energy footprint in federated systems.

\subsection{Energy Optimization}

The efforts of energy optimization on mobile devices has been revolving around software and hardware components to elongate battery lifetime, e.g., dynamic voltage and frequency scaling, resolving ``energy bugs'' from unexpected energy consumption~\cite{ebug} and coalescing packets to reduce tail energy on the wireless network interface~\cite{packet1,packet2} using the Lyapunov framework~\cite{lyaponouv}. For on-device training~\cite{wang-mm}, delegating the long-running, training workloads as a background service is a viable way to avoid interrupting normal usage. However, its performance is still unclear since the mobile systems are built with event-driven, user-centric designs to render the best performance for the foreground applications. Fortunately, the big.LITTLE architecture extends the capacity to handle concurrent low-intensity workloads on the more energy-efficient cores~\cite{arm}. Since a running foreground application would have already activated shared resources on the big cores and the background processes are typically dispatched to the little cores, co-running training with applications could take advantages of such energy disproportionality~\cite{atc}. Similar to packet coalescing~\cite{packet1,packet2}, this idea of task bundling dates back to piggyback sensing activities with applications such as web browsing and phone calls~\cite{sensys13}. However, these early works cannot be readily applied to federated learning to achieve the energy-staleness trade-off as well as bounded staleness with statistical stability. The closest works to ours are~\cite{packet1,packet2} that adopt the Lyapunov framework to achieve energy-delay trade-offs. This paper takes a step forward to consider gradient staleness induced by delayed execution and attempts to fill the gap between the machine learning algorithms and mobile systems for optimal energy efficiency.

\section{Motivation and System Model} \label{sec:motivation}

\subsection{Preliminary Experiment}
\begin{figure}[!t]
\vspace*{-0.09in}
\centering
\hspace*{-0.2in}
\begin{subfigure}[b]{0.25\textwidth}
                \includegraphics[width=1.1\textwidth]{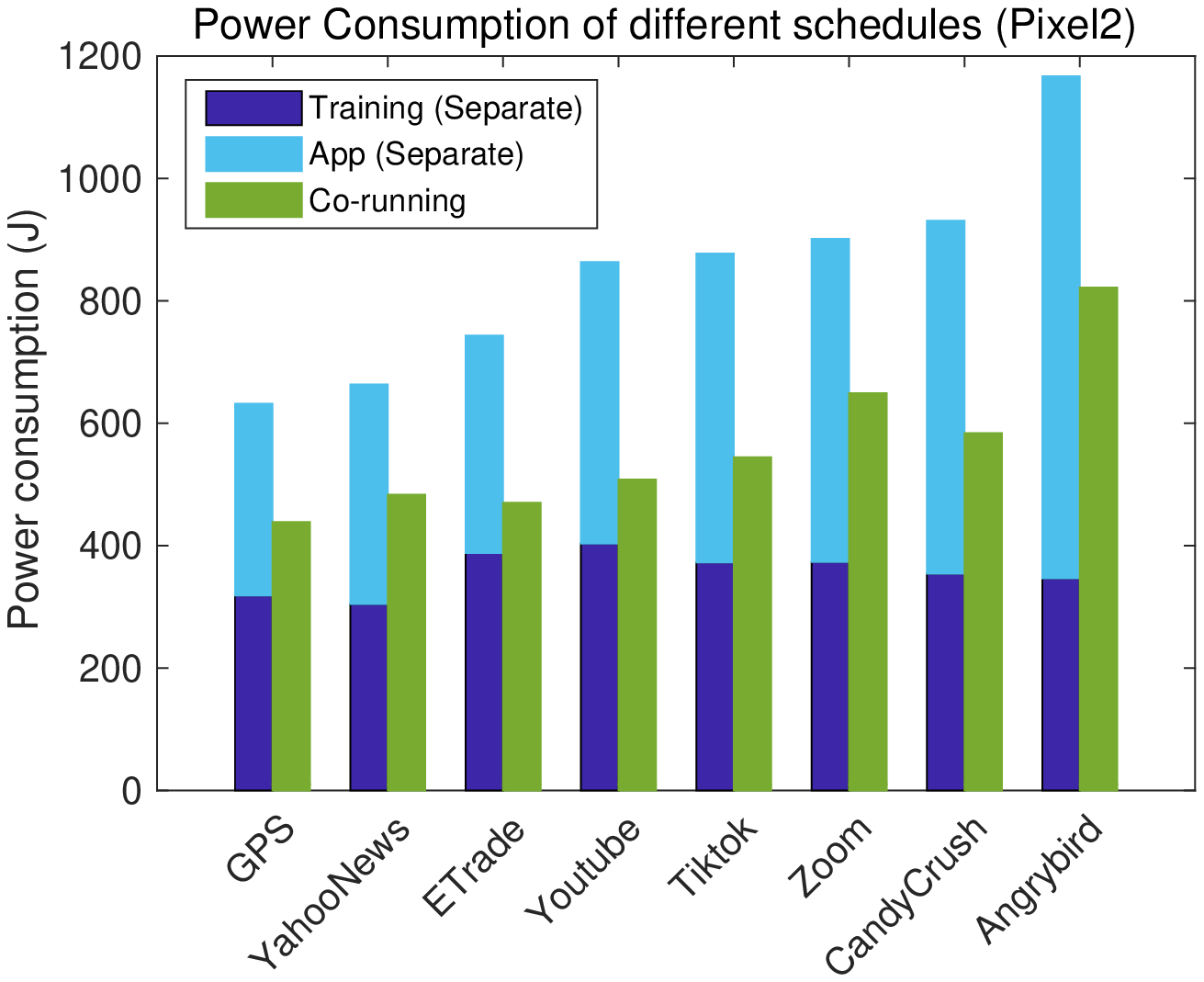}
                \vspace{-0.2in}
                \caption{}
\end{subfigure}
\begin{subfigure}[b]{0.25\textwidth}
                \includegraphics[width=1.1\textwidth]{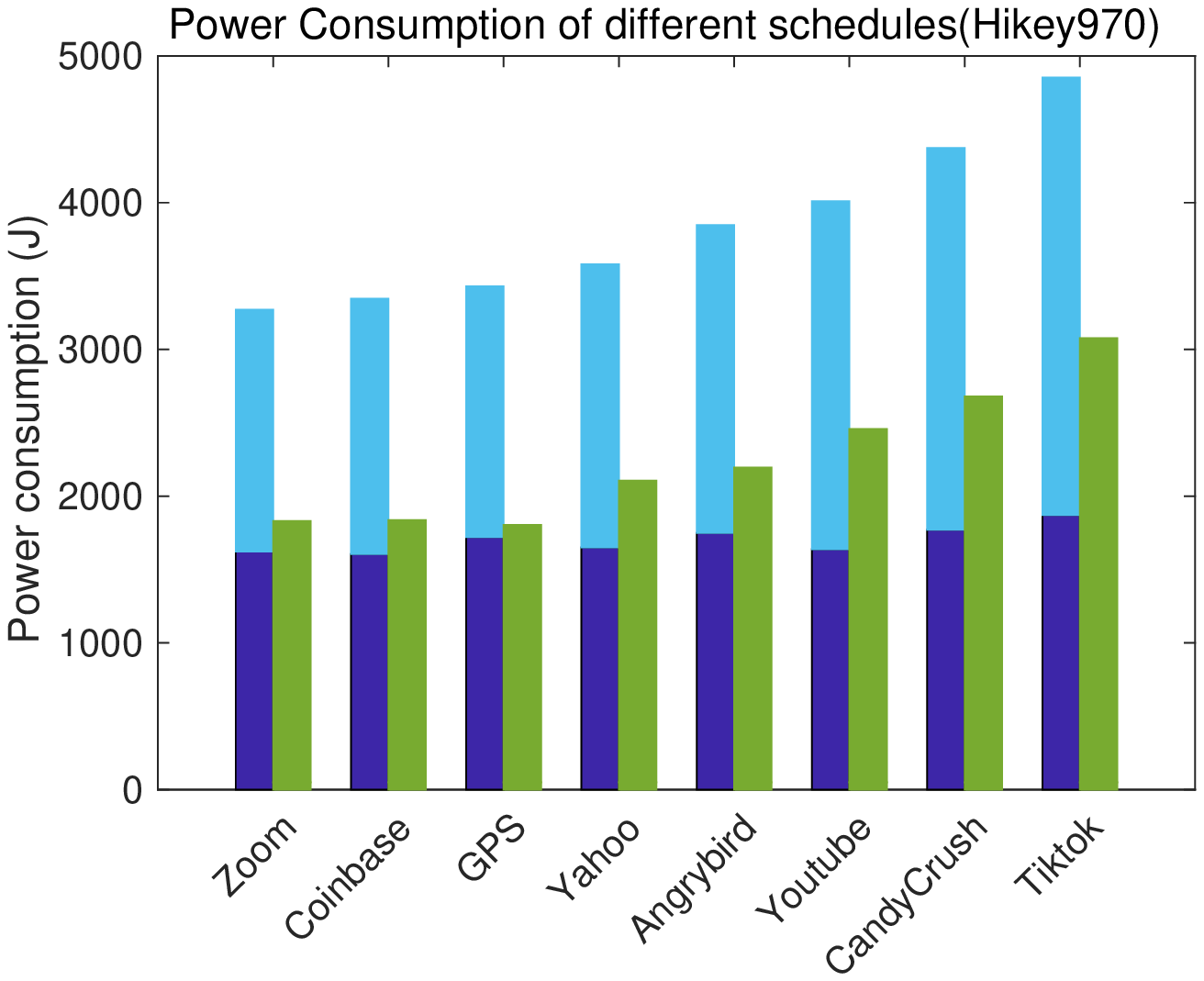}
                \vspace{-0.2in}
                \caption{}
\end{subfigure}
\vspace{-0.1in}
\caption{Power consumption of different schedules (a) Pixel2 (b) Hikey970 Dev. Board.}
\label{power_fig}
\vspace*{-0.11in}
\end{figure}

\begin{figure}[!t]
\vspace*{-0.09in}
\centering
\hspace*{-0.2in}
\begin{subfigure}[b]{0.25\textwidth}
                \includegraphics[width=1.1\textwidth]{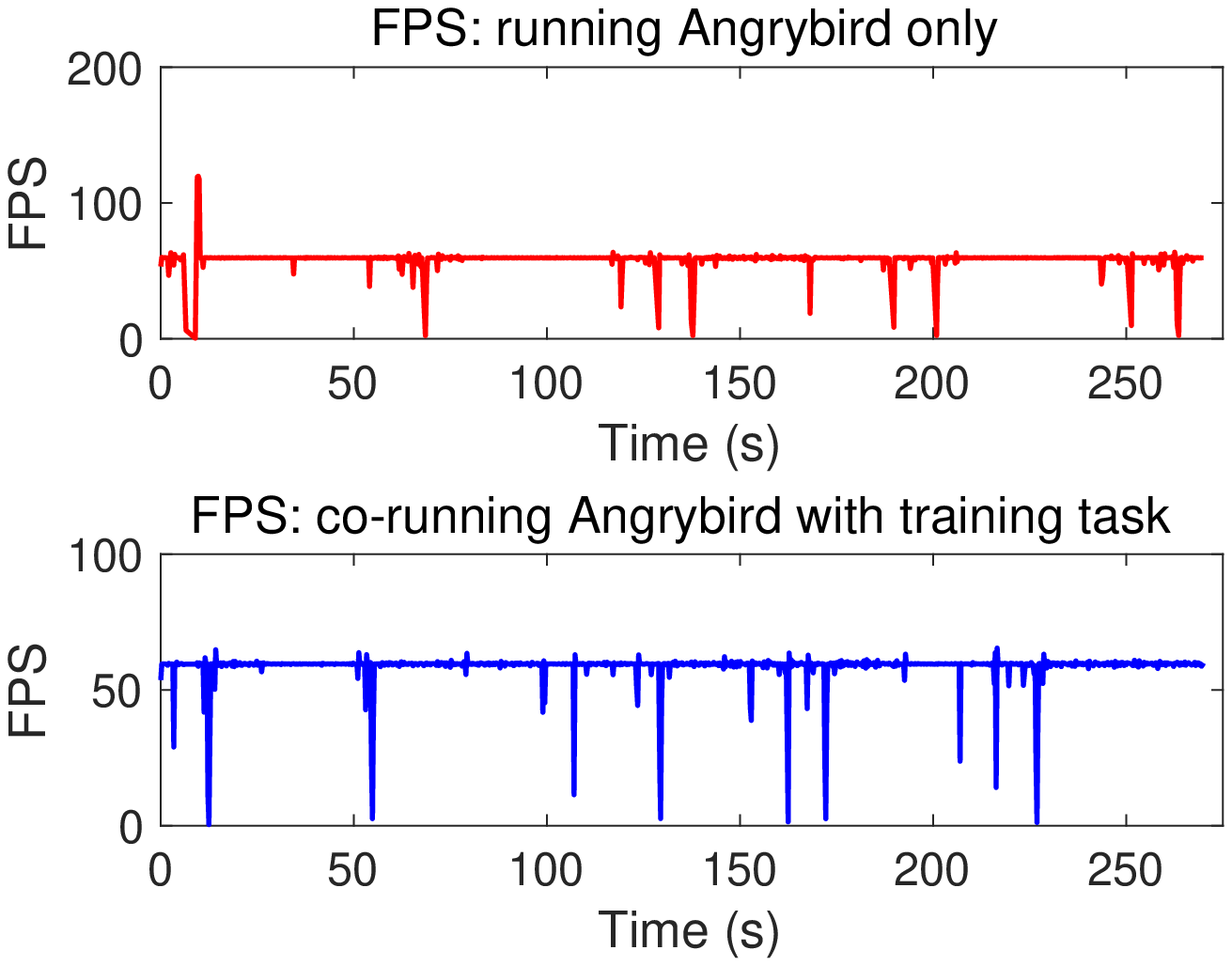}
                \vspace{-0.2in}
                \caption{}
\end{subfigure}
\begin{subfigure}[b]{0.25\textwidth}
                \includegraphics[width=1.1\textwidth]{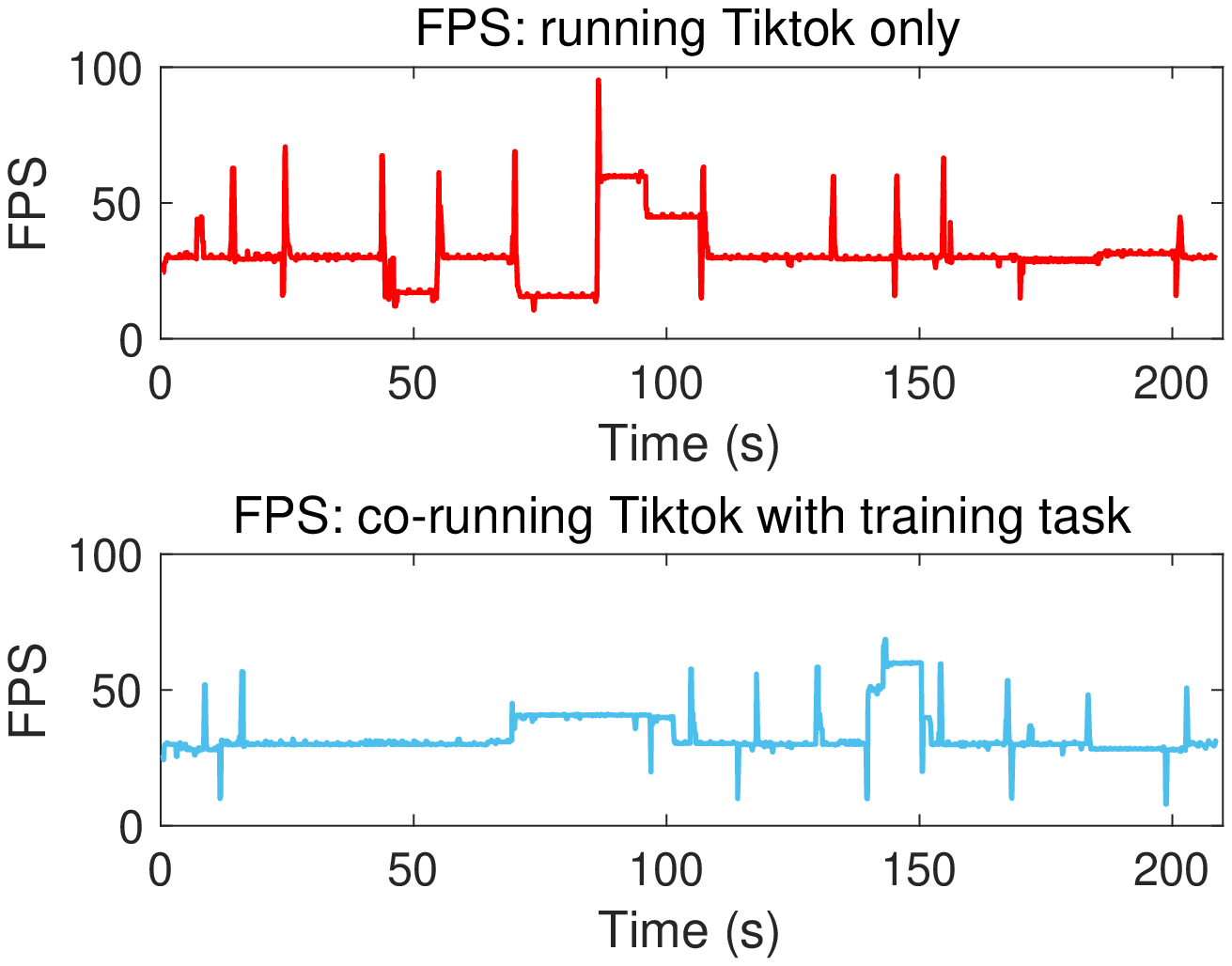}
                \vspace{-0.2in}
                \caption{}
\end{subfigure}
\vspace{-0.15in}
\caption{Performance impact measured by FPS while co-running training tasks with (a) Angrybird (b) Tiktok. }
\label{fps_fig}
\vspace{-0.15in}
\end{figure}

We motivate the design by conducting some preliminary power measurements on the HiKey 970 Development Board~\cite{hikey970} and Pixel2 smartphone (see Sec. \ref{sec:implementation} for implementation details). Assuming the user is going to run an application at a certain time, we compare the power consumption of two approaches: 1) schedule training as a service in the background, independently from the upcoming application (\emph{separate}). 2) schedule the training task to execute together with the foreground application (\emph{co-running}) and the application also stops when training finishes. Since applications have diverse resource demands and patterns of user interactions, we choose some popular applications from Google Play as shown in Fig. \ref{power_fig}. To verify that co-running does not lead to noticeable slowdown, we also perform some experiments on Pixel2 to see the rendering effects perceived by the user as measured in Frame Per Second (FPS) in Fig. \ref{fps_fig}. The important observations are summarized below.

\textbf{Observations 1}. Compared to separate scheduling, co-running offers 35-50\% energy saving. We notice that the little cores designated for executing the training task typically have 95-98\% utilizations, whereas the big cores have 30-50\% utilization depending on the foreground application.

The energy saving originates from the asymmetric CPU microarchitecture. Though the big/little core clusters have their own cache hierarchies, the memory bandwidth is shared. If the memory resource is already activated by the highly paralleled neural operations on the little cores and kept at certain power state, foreground applications executed on the big cores should not elevate the power state too much on the shared resources. Thus, co-running typically offers a substantial energy saving compared to separate executions. This is also validated by more experiments with the homogeneous cores in Nexus 6 device as resource contention on the same cluster degrades the energy saving percentage (see more details in Sec. \ref{sec:eval}).

\textbf{Observations 2}. Co-running might lead to slowdown to the training tasks depending on the intensity of the foreground applications. For some lightweight applications such as news and web browsing, the training task does not exhibit any slowdown; for intensive applications such as gaming, we notice about 10-15\% slowdown due to resource contention since a higher priority is given to the foreground applications by design. However, co-running still provides an overall energy saving despite of slightly elongated execution time.

\textbf{Observations 3}. Co-running does not have noticeable slowdown for the foreground applications, as the average FPS stays steadily around 60 and 30 frames/s shown in Fig. \ref{fps_fig}.

\subsection{System Model}   \label{sec:model}
A device pulls the current model from the parameter server when it becomes available depending on the network condition or battery energy. Training is either immediately scheduled or postponed until an application co-running opportunity. If the decision is co-running, the power consumption is $P_i^{a'}$ on the $i$-th device; otherwise, separate executions of training and application take $P_i^{b}$ and $P_i^{a}$ respectively\footnote{The power consumption of training is stable as the CPU typically stays at the maximum frequency during training. For applications, the power consumption fluctuates due to user interaction and frequency scaling. Thus, we measure the average power consumption as shown in Table \ref{energy_measurement}.}. The execution time of training is $d_i$ at the $i$-th user. For simplicity, it is assumed that the application would last for the same time duration of the training task. After the local epoch is finished, the model is pushed to the server to update the global parameters and ready to be downloaded by other participants in the following time slots. We formally define the \emph{lag} and \emph{gradient gap} to quantify gradient staleness.

\begin{figure}[!t]
\centering
\hspace*{-0.1in}
\includegraphics[width=0.51\textwidth]{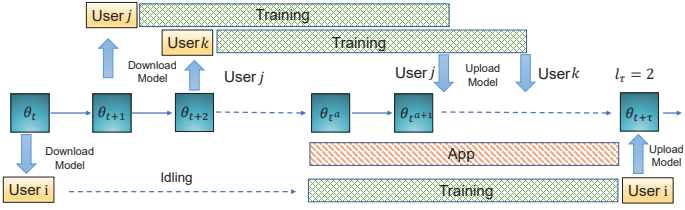}
\vspace*{-0.11in}
    \caption{Scheduling and gradient staleness in ASync-SGD.}
\label{schedule_timing_fig}
\vspace{-0.15in}
\end{figure}

\textbf{Definition 1.} (\emph{Lag}) The lag $l_\tau$ is defined as the number of updates from other users that have been made to the global model within the time interval $[t, t+\tau]$. $t$ is the initial time when the user receives the model from the server and $t+\tau$ is the time that the user finishes training and applies the parameters to the global model.

Sync-SGD guarantees the gradient aggregations are aligned in lock-step so $l_\tau=0$. For ASync-SGD, Fig.\ref{schedule_timing_fig} shows an example of three users $i,j,k$ and the control decision for $i$ is co-running at time $t^a$ when the application arrives. Users $j$ and $k$ immediately perform training without waiting for the applications and finish before $t+\tau$. Therefore, during the time interval of $[t, t+\tau]$, there are $l_\tau=2$ updates made from other users $j$ and $k$ to the global model, whereas the model update at $t+\tau$ is computed from a stale version at time $t$. Using the metric of lag alone cannot precisely quantify the difference between the two updates. Thus, we introduce another definition.

\textbf{Definition 2.} (\emph{Gradient Gap}) The gradient gap $g(t,t+\tau)$ can be calculated by the norm difference of the parameters $\theta_t$ and $\theta_{t+\tau}$~\cite{gap_1,gap_2},
\begin{equation}
\small
g(t,t+\tau) = \norm{\theta_{t+\tau} - \theta_t}_2,  \label{gradient_gap}
\end{equation}

We adopt the efficient \emph{Linear Weight Prediction}~\cite{mlsys} to estimate the global parameter $\theta_{t+\tau}$ in the future time $t+\tau$,
\begin{equation}
\small
\theta_{t+\tau} = \theta_t - \eta \frac{1-\beta^{l_\tau}}{1-\beta} v_{t}.   \label{lwp}
\end{equation}
$\eta$ is the learning rate. $\beta$ and $v_{t}$ are the momentum coefficient and vector defined in Eq. \eqref{momentum}. Plugging Eq. \eqref{lwp} into Eq. \eqref{gradient_gap}, we have
\begin{equation}
\small
g(t,t+\tau) = \norm{\eta \frac{1-\beta^{l_\tau}}{1-\beta} v_{t}}_2.   \label{lwp_gap}
\end{equation}


\section{Offline Scheduling Problem}  \label{sec:offline}
In this section, we first study an offline solution assuming that all application occurrences are known, which serves as an optimal solution and baseline for the online algorithm proposed next.

\textbf{Problem Formulation.} Energy optimization aims to achieve two conflicting goals to maximize the energy saving and avoid staleness. Given the application arrivals, the goal is to maximize the sum of energy saving from all the users: decide whether to co-run training with application for each user. Denote the number of users by $n$. For the $i$-th device, the energy saving $s_i = P_i^b + P_i^a - P_i^{a'}$ if the decision is to co-run with application (decision variable $x_i=1$); otherwise the energy saving is $0$ ($x_i=1$).
\begin{equation}
\vspace*{-0.05in}
\small
\textbf{P1:} \hspace{0.2in} \max \sum\limits_{i=1}^{n} s_i x_i   \label{obj1}
\vspace*{-0.05in}
\end{equation}
\textbf{s.t.}
\begin{eqnarray}
\vspace*{-0.05in}
\small
&& \sum\limits_{i=1}^{n} g_i(t_i,t_i+\tau_i) x_i \leq L_b, \label{constraint1_1} \\
&& x_i \in \{0,1\}.  \label{constraint1_2}
\end{eqnarray}
Constraint \eqref{constraint1_1} imposes that the sum of gradient gaps is bounded by $L_b$ and \eqref{constraint1_2} makes $x_i$ $0$-$1$ valued. This can be considered as a \emph{Knapsack Problem}~\cite{knapsack}, which maximizes the total value of items under a weight capacity and our problem maximizes the energy saving under the staleness bound $L_b$. Since the problem is NP-complete, it can be efficiently solved by utilizing the optimal sub-structure with dynamic programming. The equation of the maximal energy saving $S_i(y)$ is,
\begin{equation}
\small
S_i(y) = \begin{cases}
S_{i-1}(y), \hspace{0.1in} 0 < y \leq g_i(t_i,t_i+\tau_i)\\
\max \big\{S_{i-1}(y), S_{i-1}(y - g_i(t_i,t_i+\tau_i))+s_i \big\}, \\ g_i(t_i,t_i+\tau_i) \leq y \leq L_b. \\
\end{cases}
\end{equation}
A key difference from the original Knapsack solution is that $g_i(t_i,t_i+\tau_i)$ is computed based on the lag $l_{\tau_i}$, which in turn, depends on the decisions of other users - this creates a looping situation. We know that, in the worst case, the lag $l_{\tau_i}$ is bounded by $n-1$ because the rest of devices could have all made their updates within $\tau_i$. To tighten this, we further reduce this value as described in the next lemma. This is because given all the beginning time, application arrival and training duration, for each device, some of the rest devices should be out of the training interval and do not count towards the lag. This allows us to obtain a tighter upper bound on $l_{\tau_i}$ without knowing the control decisions in advance. As long as this upper bound is within $L_b$, we have a feasible, sub-optimal solution.

%

\emph{Lemma 1}. Given the beginning time $t_i$, application arrival time $t_i^a$ and duration of training $d_i$ for user $i$, the lag for $i$ is bounded by,
\begin{eqnarray}
\small
l_{\tau_i} &\leq& \sum_{j=1}^{n-1} \big( \mathds{1}( t_j^a + d_j \in [t_i, t_i+ d_i] \vee [t_i^a, t_i^a + d_i]) \nonumber \\
&\vee& \mathds{1}( t_j+d_j \in [t_i, t_i+ d_i] \vee [t_i^a, t_i^a + d_i])\big),  \label{lemma1_eq}
\end{eqnarray}
in which $\vee$ denotes the logical ``or'' of the two time intervals and $\mathds{1}(\cdot)$ is one if the training ends in one of the time intervals.

\begin{proof} It can be proved by considering all possible decisions for each pair of $i$ and $j=\{1,\cdots,n-1\}$. For $i$, it has two scheduling possibilities: 1) execute training at $t_i$ and end at $t_i+d_i$ (the interval of $[t_i, t_i+ d_i]$); 2) co-run training with application arrival at $t_i^a$ and end at $t_i^a + d_i$ (the interval of $[t_i^a, t_i^a+ d_i]$). Meanwhile, any other $j$ has the similar possibilities to end at $t_j+d_j$ or $t_j^a + d_j$. If any of these intervals from $i$ and $j$ has overlaps, the gradient gap is increased by one and summed over all $n-1$ devices.
\end{proof}

Based on \emph{Lemma 1}, the offline solution is summarized in Algorithm \ref{offline_scheduling_algo} with a time complexity of $\mathcal{O}(n L_b)$, and will serve as a baseline for the online algorithm discussed next. 
\begin{algorithm}[t!]
\small
\caption{Offline Algorithm}
\label{offline_scheduling_algo}
\textbf{Input:} app. arrival time $t_i$, training execution $d_i$ $\forall i$, zero-valued matrix $S$ of size $n\times L_b$ \\
\textbf{Output:}  scheduling decisions $x_i$ and maximum energy saving.\\
Initialize $S_0(y)=0, y \geq 0$. \\
\For{$i=1\;\text{to}\;n$}
{
   \For{$j=1\;\text{to}\;L_b$}
   {
        Estimate $g_j(t_j,t_j+\tau_j)$ with Eq. \eqref{lemma1_eq} and Eq. \eqref{lwp_gap}.\\
        \If{$y \leq g_i(t_i,t_i+\tau_i)$}{
        $S_{i}(y) \leftarrow S_{i-1}(y)$.
        }
        \Else{
        $S_{i}(y) \leftarrow \max \big\{S_{i-1}(y), S_{i-1}(y - g_i(t_i,t_i+\tau_i))+s_i \big\}$.
        }
   }
}
\end{algorithm}

\section{Online Scheduling} \label{sec:online}

\begin{table*}[ht]
\centering
\small
\caption{List of important notations of device $i$.}
\begin{tabular}[t]{l l }
\hline
Notation &Definition\\
\hline
$P_i^{a'}, P_i^{d}$  &Average power consumption of training/application \emph{co-running}, and \emph{idling}.    \\
$P_i^{b}$, $P_i^{a}$ &Average power consumption of separate executions of \emph{training} and \emph{application}.  \\
$g_i(t,t+\tau), G(t,t+\tau)$  &Gradient gap between time $t$ and $t+\tau$ and the sum of gradient gaps from all the devices.    \\
$\alpha(t), s(t)$  &Control decision \{``schedule'', ``idle''\}, application status \{``app'', ``no app''\}. \\
\hline
\end{tabular} \label{notations}
\vspace{-0.15in}
\end{table*}%

Offline scheduling assumes the future application arrival as a priori. In this section, we propose an online scheduling with the Lyapunov framework that only relies on the current observation. We consider a task queue for the entire system as defined below.


\textbf{Definition 3.} (\emph{Queue Dynamics}). The task queue represents the number of users waiting to be scheduled. The arrival of users can be considered as a random process $A(t)$. The queue backlog will increase by $A(t)=n$ if a number of $n$ users are ready to start training at $t$. If $m$ users finish their training in a time slot, the queue backlog is reduced by $b(t) = m$.

Assume time is equally slotted with the length of $t_d$. The system makes a control decision $\alpha(t)$ at time $t$. The energy consumption $P_i(t)$ of the $i$-th device depends on how training is scheduled and the current application status $s(t) = \{\mlq app \mrq, \mlq no \hspace{0.05in} app \mrq \}$, i.e., $P_i(t)= P_i(\alpha(t), s(t))$:
\begin{equation}
\small
P_i(t) = \begin{cases}
P_i^{a'} t_d, \hspace{0.1in} \alpha(t)= \mlq schedule \mrq, s(t) = \mlq app \mrq\\
P_i^b t_d, \hspace{0.1in} \alpha(t) = \mlq schedule \mrq, s(t) = \mlq no \hspace{0.05in} app \mrq\\
P_i^a t_d, \hspace{0.1in} \alpha(t)=\mlq idle \mrq, s(t) = \mlq app \mrq\\
P_i^d t_d, \hspace{0.1in} \alpha(t) = \mlq idle \mrq, s(t) = \mlq no \hspace{0.05in} app \mrq. \\
\end{cases}\label{power_eq}
\end{equation}
According to the experimental measurements, $P_i^{a'} > P_i^{a} > P_i^{b} > P_i^{d}$. The corresponding service rate for $i$ is\footnote{To simplify the analysis, we take an approximation here to make the actual service and deduction of queue length at $t+d_i$ to be effective at $t$.},
\begin{equation}
\small
b_i(t) = \begin{cases}
1, \hspace{0.1in} \alpha(t) = \mlq schedule \mrq\\
0. \hspace{0.1in} \alpha(t) = \mlq idle \mrq \\
\end{cases}
\end{equation}
and the service rate in the system is $b(t) = \sum\limits_{i=0}^{n}b_i(t)$. The gradient gap is,
\begin{equation}
\small
g_i(t,t+\tau_i) = \begin{cases}
\norm{\eta \frac{1-\beta^{l_{d_i}}}{1-\beta} v_{t}}_2, \hspace{0.1in} \alpha(t) = \mlq schedule \mrq\\
g_i(t-1,t+\tau_i-1) + \epsilon. \hspace{0.1in}  \alpha(t) = \mlq idle \mrq \\
\end{cases}\label{gradient_gap_scheduling}
\end{equation}
If the decision is to schedule training, the gap is computed using Eq. \eqref{lwp_gap} with lag $l_{d_i}$ during the execution time of $d_i$; if the decision is to remain idle, the gap is cumulative from the previous slot plus a small time-averaged gap increment $\epsilon$, which estimates the impact on the gradient gap for each idling time slot. The sum of gradient gaps is $G(t,t+\tau) = \sum\limits_{i=0}^{n} g_i(t,t+\tau)$.

\textbf{Problem Formulation.} Our goal is to minimize the time-averaged energy consumption of training tasks in the system of $n$ users,
\begin{equation}
\small
\textbf{P2:} \hspace{0.2in} \limsup_{T \rightarrow \infty} \frac{1}{T} \sum_{t=1}^{T} \sum_{i=1}^{n} \mathbb{E}\{P_i(t)\}
\end{equation}
\textbf{s.t.}
\begin{eqnarray}
\small
& \limsup\limits_{T \rightarrow \infty} \frac{1}{T} \sum\limits_{t=1}^{T} \sum\limits_{i=1}^{n} g_i(t,t+\tau) \leq L_b  \label{constraint21}
\end{eqnarray}
Eq. \eqref{constraint21} guarantees that the sum of gradient gaps from all the participants is bounded in a time averaged sense. \textbf{P2} can be transformed into the queue stability problem under the Lyapunov optimization framework. Given the arrival rate $A(t)$ and service rate $b(t)$, the queueing dynamics is,
\begin{equation}
\small
Q(t+1) = \max \big(Q(t) - b(t), 0\big) + A(t) \label{queueing_eq1}
\end{equation}
with the initial $Q(0) = 0$. We define a \emph{virtual queue} $H(t)$ for constraint \eqref{constraint21},
\begin{equation}
\small
H(t+1) = \max \big(H(t) + \sum\limits_{i=1}^{n} g_i(t,t+\tau) - L_b, 0 \big) \label{queueing_eq2}
\end{equation}
and the initial $H(0)=0$. We concatenate the actual and virtual queues into $\Theta(t) = [\mathbf{Q}(t), \mathbf{H}(t)]$, define the Lyapunov function $L(\Theta(t))$ as the queue congestion of the backlogged training tasks,
\begin{equation}
\small
L(\Theta(t)) \overset{\Delta}{=} \frac{1}{2} (Q(t)^2 + H(t)^2),  \label{queueing_all}
\end{equation}
and the Lyapunov drift function $\Delta (\Theta(t))$ as:
\begin{equation}
\small
\Delta (\Theta(t)) \overset{\Delta}{=} \mathbb{E}\{ L(\Theta(t+1)) - L(\Theta(t))|\Theta(t)\} \label{drift_eq}
\end{equation}
It represents the change in the Lyapunov function in time slot $t$ representing the scalar volume of queue congestions. The new optimization problem is to minimize the drift-plus-penalty:
\begin{equation}
\small
\textbf{P3:} \hspace{0.25in} \min \Delta (\Theta(t)) + V \mathbb{E}\{P(t)| \Theta(t)\}  \label{new_obj}
\end{equation}
$V$ is the control parameter to balance between energy and staleness. Following the Lyapunov framework, the key is to obtain the upper bound of the drift as described in the following Lemma.

\emph{Lemma 2}: Given the queue backlogs $Q(t)$, arrival rates $A(t)$ and service rate $b(t)$ and the gradient gaps, we have the following upper bound for the drift-plus penalty term,
\begin{fleqn}
\begin{equation}
\small
\begin{aligned}[b]
& \Delta (\Theta(t)) + V \mathbb{E}\{P(t)| \Theta(t)\}  \leq B + V \mathbb{E}\{P(t)| \Theta(t)\} + \\
& Q(t)\mathbb{E}\{(A(t) - b(t)| \Theta(t)\} + H(t) \mathbb{E}\{G(t, t+\tau)-L_b | \Theta(t)\} \label{lemma2}
\end{aligned}
\end{equation}
\end{fleqn}
where $B = \frac{1}{2} (A_{max}^2 + B_{max}^2 + G^2_{max} + L_b^2)$ is a positive constant. The proof can be found in Appendix \ref{appendix}.

Our algorithm observes the current queue backlogs of $Q(t), H(t)$ and application status $s(t)$ to make a decision of $\alpha(t) \overset{\Delta}{=}\{ \mlq schedule \mrq ,\mlq idle \mrq \}$ that minimizes the R.H.S. of the drift bound Eq. \eqref{lemma2}, which is equivalent to the objective in Eq. \eqref{new_obj}.
\begin{equation}
\small
\min \left( V \sum\limits_{i=1}^{n} P_i(t) - Q(t) \sum\limits_{i=1}^{n} b_i(t) + H(t) \sum\limits_{i=1}^{n} g_i(t, t+\tau_i)\right) \label{rhs_1}
\end{equation}
This formulation makes online decisions based on the current observations and does not need a-priori knowledge of the arrival rates. With the information of application usage, a centralized implementation can be conducted in $\mathcal{O}(n)$ at the parameter server. However, since application usage are considered as private and their patterns can be used to re-identify specific users~\cite{app_privacy}, centralization carries certain privacy risks.

\subsection{Distributed Implementation}
The minimization of Eq. \eqref{rhs_1} can be achieved in a distributed manner via appropriate information exchange between the server and the users. We design distributed implementations into the Lyapunov framework that can mitigate the privacy leakage of application usage to the parameter server. In time $t$, each user minimizes Eq. \eqref{rhs_1} from the control decision space based on status of application usage and queue backlogs, thus the application usage $s_i(t)$ is not leaked to the server. In the last term of Eq. \eqref{rhs_1}, the user computes the gradient gap $g_i(t_i,t_i +\tau_i)$ according to Eq. \eqref{gradient_gap}. If the decision is ``schedule'', the number of updates in the time interval of $[t, t+ d_i]$ can be supplied by the server with the estimated arrival time of the running tasks; otherwise, the gap accumulates from the previous value plus a small increment according to Eq. \eqref{gradient_gap_scheduling}. Hence, for each user $i$, the decision making fully depends on its own status except the lag value supplied from the server, which reveals little information about application usage compared to the centralized implementation. The procedures are summarized in Algorithm \ref{online_scheduling_algo} with a computational complexity of $\mathcal{O}(1)$ at each user and communication overhead of $\mathcal{O}(n)$ at the server.

\begin{algorithm}[t!]
\caption{Distributed Online Scheduling Algorithm}
\label{online_scheduling_algo}
\textbf{Input:} Queue backlogs $Q(t)$ and $H(t)$, control parameter $V$, and action space $\Omega$, learning rate $\eta$, momentum vector $v$. \\
\textbf{Output:} Scheduling decisions $\forall i$. \\
\For{$i=1\;\text{to}\;n$}
{
Send duration $d_i$ to the server, and receives lag $l_{d_i}$ from the server.\\
Estimate $g_i(t,t+\tau_i)$ with Eq. \eqref{lwp_gap}.\\
$\alpha_i(t) \leftarrow \argmin\limits_{P_i, b_i, g_i} V P_i(t) - Q(t) b_i(t) + H(t) g_i(t, t+\tau_i).$\\
Inform control decision $\alpha_i(t)$ to server. \\
}
Server: Update $Q(t), H(t)$ according to Eqs. \eqref{queueing_eq1} and \eqref{queueing_eq2} respectively according to $\alpha(t)$.
\end{algorithm}

\vspace{-0.1in}
\subsection{Illustration of Control Decisions}
In the objective of Eq. \eqref{rhs_1}, $H(t) \sum\limits_{i=1}^{n} g_i(t, t+\tau)$ can be viewed as a penalty term when there are backlogs in the virtual queue. When there is no backlog ($Q(t), H(t)=0$), we only have the first term in Eq. \eqref{rhs_1} so the control decision is to always set the device to idle. This matches with the intuition to wait for better co-running opportunities.

\textbf{No Staleness from the Virtual Queue.} There could be cases that there are queue backlogs in $Q(t)$, but for the virtual queue $H(t)$, the cumulative gradient gap has not exceeded the bound $L_b$, i.e., $H(t-1)+\sum_{i=1}^{n} g_i(t-1,t+\tau-1) \leq L_b$. Hence, $H(t)\sum_{i=1}^{n} g_i(t,t+\tau) = 0$ and we derive the decision of,
\begin{equation}
\small
\alpha_i(t) = \argmin\limits_{\alpha_i(t)} \begin{cases}
(V P_i^{a'} t - Q(t), V P_i^a t), \hspace{0.03in} s(t) = \mlq app \mrq   \\
(V P_i^b t - Q(t), V P_i^d), \hspace{0.03in} s(t) = \mlq no \hspace{0.05in} app \mrq   \\
\end{cases}\label{condition1}
\end{equation}
The decision can be made by simply observing $Q(t)$: for $s(t) = \mlq app \mrq$, the decision is to co-run if $Q(t) \geq V t (P_i^{a'} - P_i^a)$; otherwise, the decision is idling. Similarly, for $s(t) = \mlq no \hspace{0.05in} app \mrq$, the decision is to execute as a background process when $Q(t) \geq V t (P_i^{b} - P_i^d)$ or set to idle otherwise. As a result, the controller would wait until the queue length reaches a certain level.

\textbf{With Gradient Staleness.} When $H(t) g(t)>0$, the penalty term $H(t) g(t)$ is active so the control decision accounts for possible staleness.
\begin{equation}
\small
\hspace{-0.02in} \alpha_i(t) = \argmin\limits_{\alpha_i(t)}\begin{cases}
\bigg( V P_i^{a'} t - Q(t)+H(t) \norm{\eta \frac{1-\beta^{l_\tau}}{1-\beta} v_{t}}_2, V P_i^a +
\\H(t)\big(g_i(t-1,t+\tau-1) + \epsilon\big) \bigg), \hspace{0.02in} s(t) = \mlq app \mrq   \\
\vspace{0.08in}
\bigg(V P_i^b t - Q(t) + H(t) \norm{\eta \frac{1-\beta^{l_\tau}}{1-\beta} v_{t}}_2, V P_i^d+\\
H(t)\big(g_i(t-1,t+\tau-1) + \epsilon\big) \bigg), \hspace{0.02in} s(t) = \mlq no \hspace{0.03in} app \mrq   \\
\end{cases}\label{condition2}
\end{equation}
The relevant control decisions can be made by observing $Q(t), H(t)$ and compute the rest of the values in Eq. \eqref{condition2}.

\subsection{Optimality Analysis}
The optimality of the problem is derived in \emph{Theorem 1}.

\emph{Theorem 1}. Let $L(\Theta(t))$ defined by Eq. \eqref{queueing_all} and $L(\Theta(0)) = 0$. $P^\ast$ is the optimal power consumption. For constants $B,V \geq 0$, the queues of $\Theta(t)$ are mean rate stable and the time-averaged power consumption and queue backlogs are bounded by:
\begin{equation}
\small
\limsup\limits_{T \rightarrow \infty} \frac{1}{T} \sum\limits_{t=0}^{T-1} \mathbb{E}\{P(t)\}  \leq  \frac{B}{V} + P^\ast \label{theorem1_1}
\end{equation}
\begin{equation}
\small
\limsup\limits_{T \rightarrow \infty} \frac{1}{T} \sum\limits_{t=0}^{T-1} \mathbb{E}\{\Theta(t)\}  \leq  \frac{B}{\epsilon_1} + \frac{V (P^\ast - \overline{P})}{\epsilon_1} \label{theorem1_2}
\end{equation}

The proofs are detailed in Appendix \ref{appendix-2}. The performance bounds Eqs. \eqref{theorem1_1}, \eqref{theorem1_2} demonstrate an $[\mathcal{O}(1/V), \mathcal{O}(V)]$ energy-staleness trade-off: by arbitrarily increasing $V$, we can make $\frac{B}{V} \rightarrow 0$ and the time-averaged power consumption close to the optimal value, whereas the staleness grows linearly with $V$.

\section{System Implementation}  \label{sec:implementation}

\begin{table*}[!ht]
\hspace*{-0.1in}
\centering
\small
\begin{tabular}{l@{\hspace{0.1in}} c @{\hspace{0.1in}} c @{\hspace{0.1in}} c @{\hspace{0.1in}} c@{\hspace{0.1in}}  c @{\hspace{0.1in}} c @{\hspace{0.1in}}c @{\hspace{0.1in}}c @{\hspace{0.1in}}
c @{\hspace{0.1in}} c @{\hspace{0.1in}} c @{\hspace{0.1in}} c @{\hspace{0.1in}} c @{\hspace{0.1in}} c @{\hspace{0.1in}} c @{\hspace{0.1in}} c}
\toprule
 & \multicolumn{4}{c}{Nexus6} & \multicolumn{4}{c}{Nexus6P} & \multicolumn{4}{c}{Hikey970} & \multicolumn{4}{c}{Pixel2}\\
\cmidrule(lr){2-5} \cmidrule(lr){6-9} \cmidrule(lr){10-13} \cmidrule(lr){14-17}
Apps      &app &co-run &time &saving(\%)    &app &co-run &time &saving(\%)     &app &co-run &time &saving(\%)      &app &co-run &time &saving(\%)   \\
\midrule
Training    & 1.8 & -- &204s  &--      & 0.9  & --  & 211s & --   & 7.87  & -- & 213s & --    & 1.35 & -- & 223s & --\\
Map  & 3.4 & 3.5  & 274s & 26\%             & 0.5  & 1.3 & 225s & 3\%       & 8.82  &9.42  &186s & 47\%         & 1.60 & 2.20 & 196s & 30\%   \\
News  & 1.7 & 2.2 & 239s & 32\%            & 0.44  & 1.2 & 362s & -24\%        & 9.17  & 9.76  & 210s & 43\%        & 1.82 & 2.40 & 197s & 28\%   \\
Etrade    & 1.4 & 2.4  & 236s & 17\%       & 0.48  & 0.96 & 228s & 27\%    & 8.50  &9.15  &195s & 47\%   & 1.72 & 2.23 & 206s & 30\%    \\
Youtube   & 0.5 & 1.9 & 284s & -4\%        & 0.53  & 1.2 & 220s & 14\%    & 9.15  & 11.45  & 210s & 33\%   & 2.04 & 2.21 & 226s & 35\%     \\
Tiktok  & 1.6 & 2.3 & 296s & 18\%     & 1.0  & 1.1 & 675s & 14\%    & 11.0  & 11.2 & 271s & 35\%    & 2.37 & 2.52 & 212s & 34\%   \\
Zoom    & 1.2 & 2.1  & 370s & 4\%      & 1.4  & 1.6 & 340s & 18\%         & 7.89  & 8.53 & 209s & 46\%   & 2.57 & 3.11 & 206s & 23\%   \\
CandyCru  & 1.3 & 2.3 & 997s & -39\%     & 0.7  & 1.3 & 280s & 9\%   & 11.1  & 11.26 & 233s & 38\%    & 2.89 & 2.92 & 199s & 34\%  \\
Angrybird    & 2.5 & 2.8 & 400s & 18\%     & 1.1  & 1.2 & 620s & 15\%    & 10.1  & 10.7 & 200s & 42\%    & 2.86 & 2.88 & 285s & 26\%   \\
\bottomrule
\end{tabular}
\vspace{-0.05in}
\caption{Averaged energy measurements - battery power (W) and execution time (s) running LeNet-5 of CIFAR10 dataset.}  \label{energy_measurement}
\end{table*}

To conduct neural net training on Android, we adopt a Java-based Deep Learning framework called DL4J~\cite{dl4j}, which provides seamless integration with the Android OS. The backend neural computations are supported by OpenBLAS cross-compiled for the ARM platforms. We pre-load the CIFAR10 dataset~\cite{cifar10} into the flash storage of the phone and retrieve in batch size of $20$. The Training App is implemented using the Android \texttt{JobScheduler} framework designed for long-running operations in the background without interfering with the foreground applications. Conditions such as networking connectivity (Wifi/4G), device status (idling or charging) and execution time window can be specified to offer fine-grained control. Once the job scheduler is started using \texttt{onStart}, a new thread is created to initialize the neural network in the device's memory. We enable the \texttt{largeHeap} to give the App 512MB memory to avoid memory errors. The number of CPU cores designated for background services is specified by the vendor, which can be found in \texttt{/dev/cpuset/background/cpus}. E.g., Pixel2 utilizes the two little cores and Nexus6P, Hikey970 only run on the one little core and the rest of the three little cores are reserved for system background process. Note that the default kernel (e.g., CPU affinity, priority, frequency scaling) is used and no root access is required throughout the paper. We set the number of training threads to $2$ or $1$ according to the vendor specifics, because a large value would conversely lead to potential contentions to keep cache coherence and ultimately slow down computations.

The Android kernel might kill the background training process to save memory and optimize battery lifetime, particularly when the neural network involves intensive computations. We do not find the service being killed while running LeNet-5, but introducing more convolutional layers with large filter size would invoke the automatic background limitation because those layers are the major resource consumers. In practice, there also exists a few ``diehard'' tricks such as escalating the app priority, service binding, etc~\cite{diehard}. We intend to incorporate some of these methods in the future, whereas a fundamental solution to this problem from the Mobile OS is out of the scope of this paper.

The communication part is handled by the Retrofit Framework from Sqaure~\cite{retrofit}, which easily packages asynchronous HTTP requests to the Python-based HTTP server. For Async-SGD, once a device completes a local epoch, it creates a Retrofit \texttt{FileuploadService} to upload the local model of 2.5 MB with meta information (device ID, round \#) to the server. The server replaces the current copy of the global model upon receiving it. When the device becomes available, it downloads the current model using the \texttt{FileDownloadService} as a starting point for the next local epoch.

\begin{figure*}[ht]
\centering
\hspace*{-0.2in}
\begin{subfigure}[b]{0.24\textwidth}
                \includegraphics[width=1.02\textwidth]{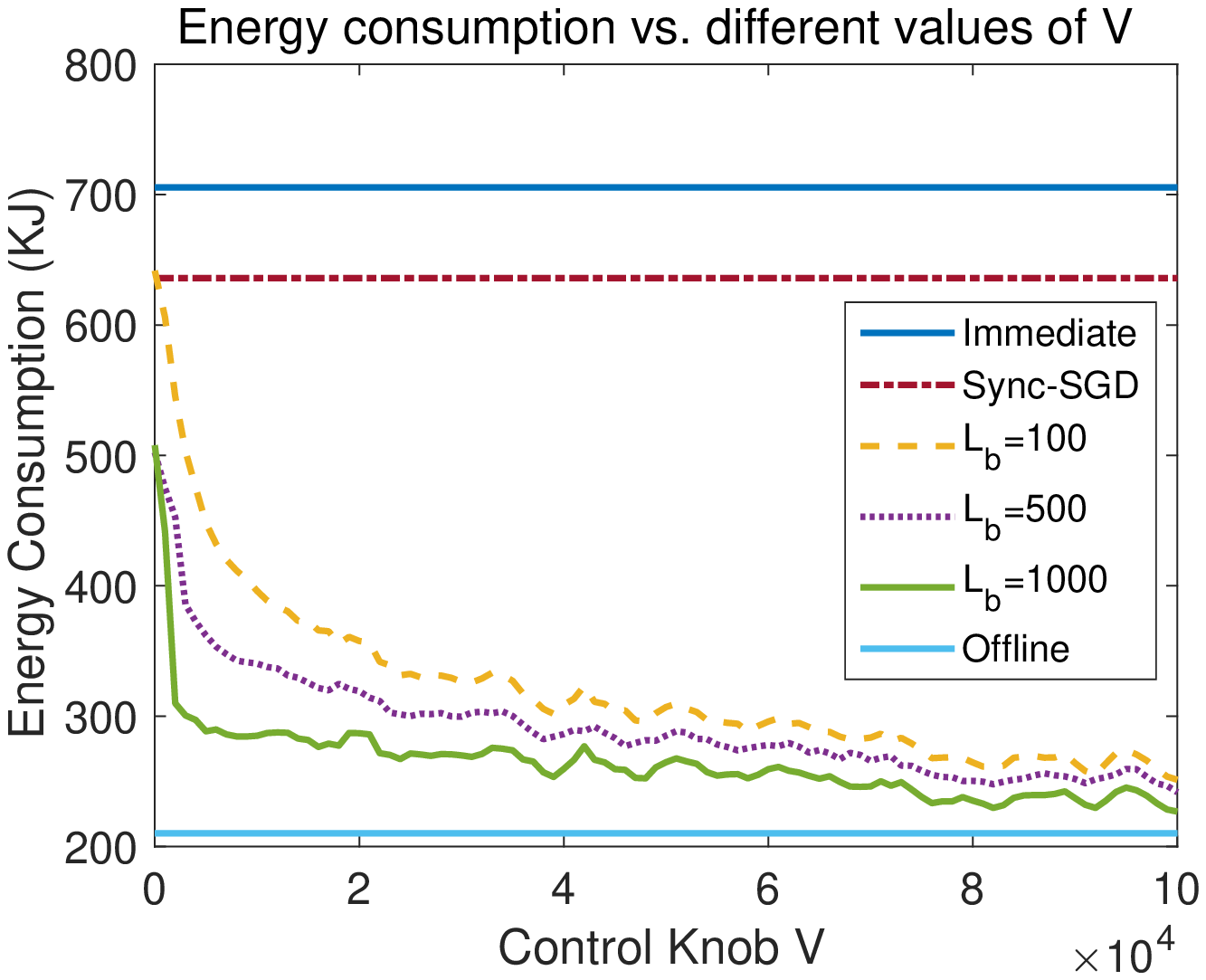}
                \vspace{-0.21in}
                \caption{}
\end{subfigure}
\begin{subfigure}[b]{0.24\textwidth}
                \includegraphics[width=1.02\textwidth]{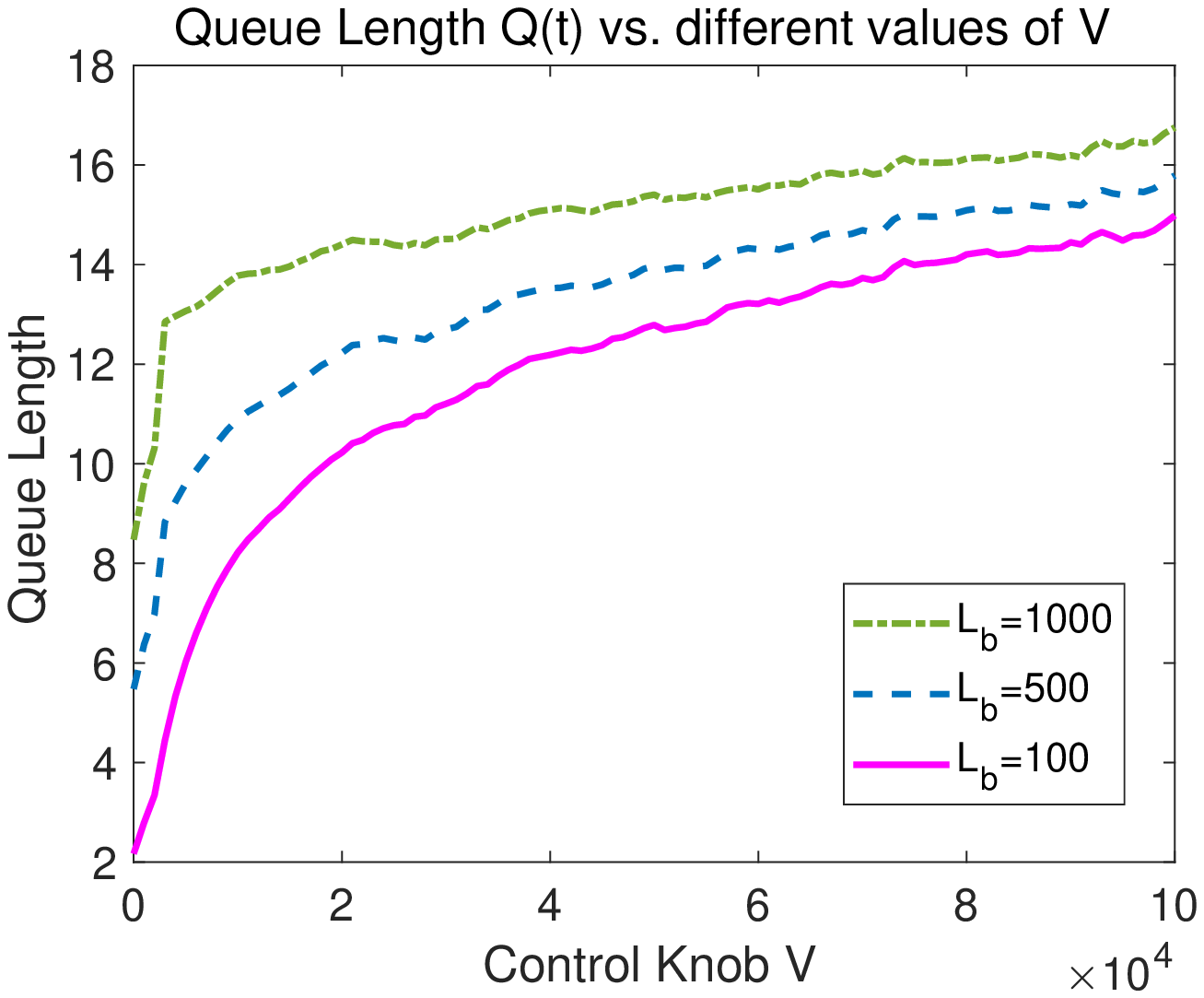}
                \vspace{-0.21in}
                \caption{}
\end{subfigure}
\begin{subfigure}[b]{0.24\textwidth}
                \includegraphics[width=1.02\textwidth]{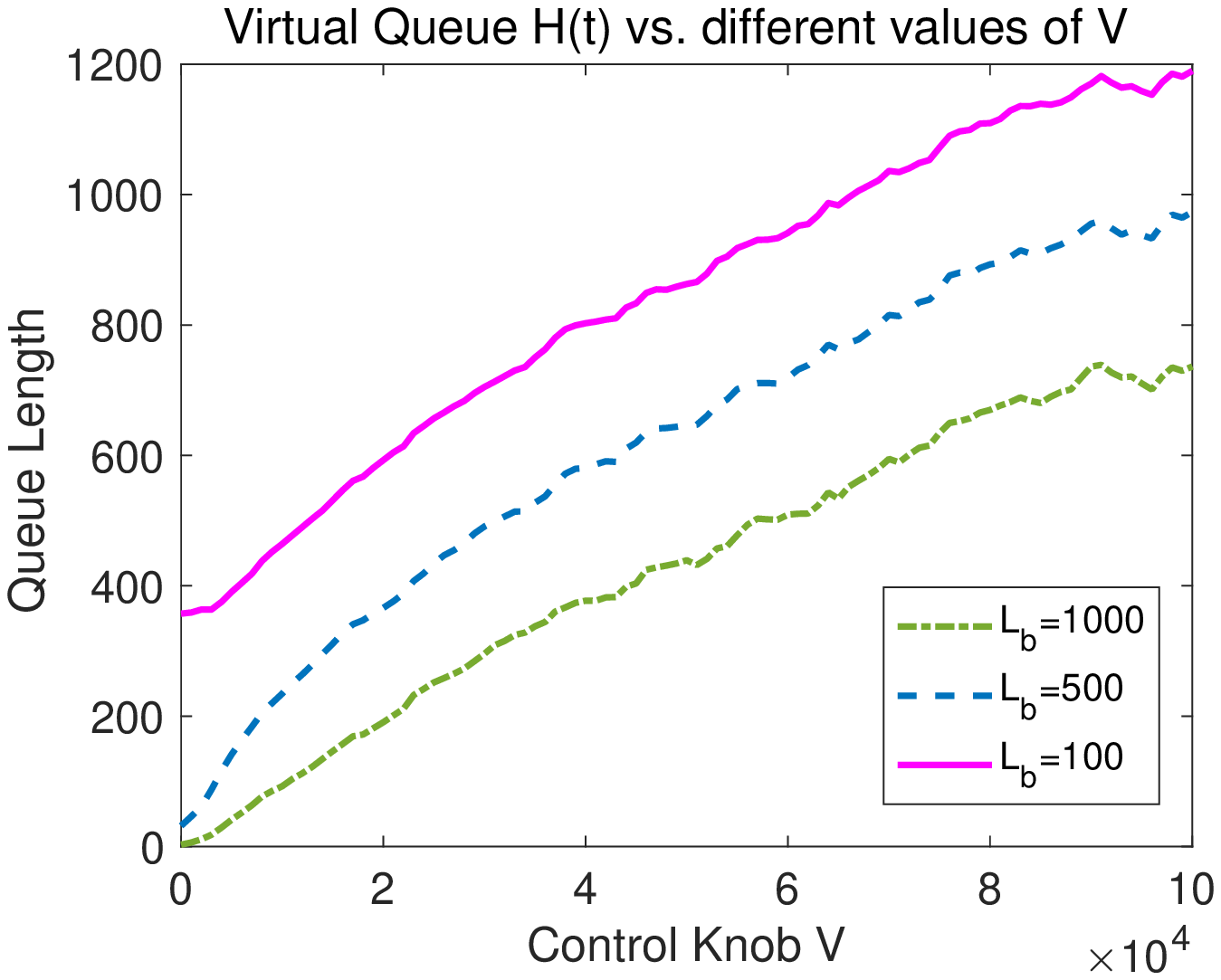}
                \vspace{-0.21in}
                \caption{}
\end{subfigure}
\begin{subfigure}[b]{0.24\textwidth}
                \includegraphics[width=1.02\textwidth]{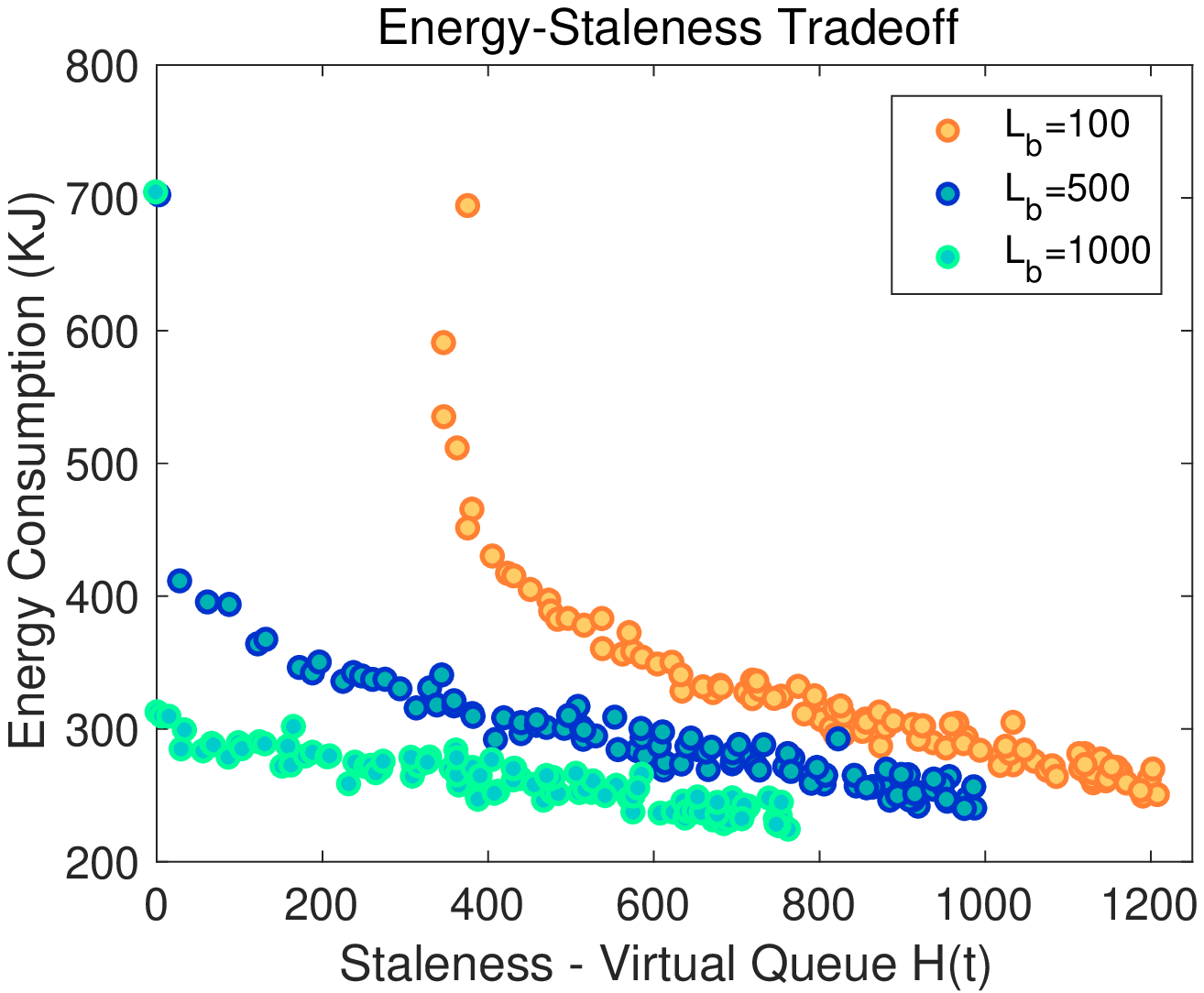}
                \vspace{-0.21in}
                \caption{}
\end{subfigure}%
\hspace*{-0.01in}
\caption{Energy consumption and trade-offs (a) Energy consumption vs. $V$; (b) Queue length $Q(t)$ vs. $V$; (c) Virtual queue length $H(t)$ vs. $V$; (d) Energy-staleness trade-off with different staleness bound $L_b$.}
\label{energy_consumption}
\end{figure*}

\section{Evaluation} \label{sec:eval}
\textbf{Testbed/Parameter Settings.} The evaluation is conducted across a dynamic set of mobile devices from different vendors: Nexus 6/6P, HiKey970 Dev. Board and Pixel2.

\begin{figure*}[!ht]
\centering
\hspace*{-0.2in}
\begin{subfigure}[b]{0.245\textwidth}
                \includegraphics[width=1.02\textwidth]{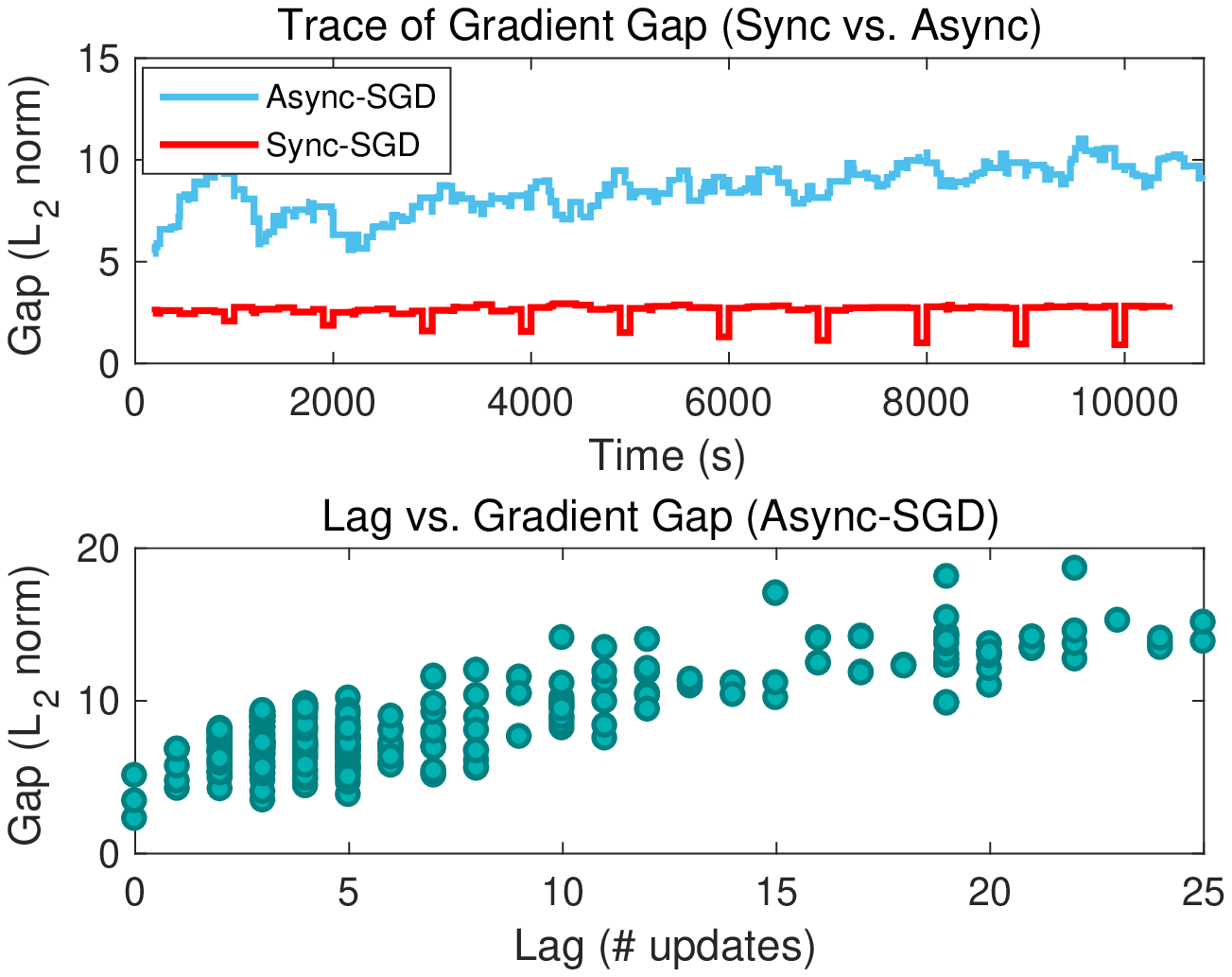}
                \vspace{-0.21in}
                \caption{}
\end{subfigure}
\begin{subfigure}[b]{0.245\textwidth}
                \includegraphics[width=1.02\textwidth]{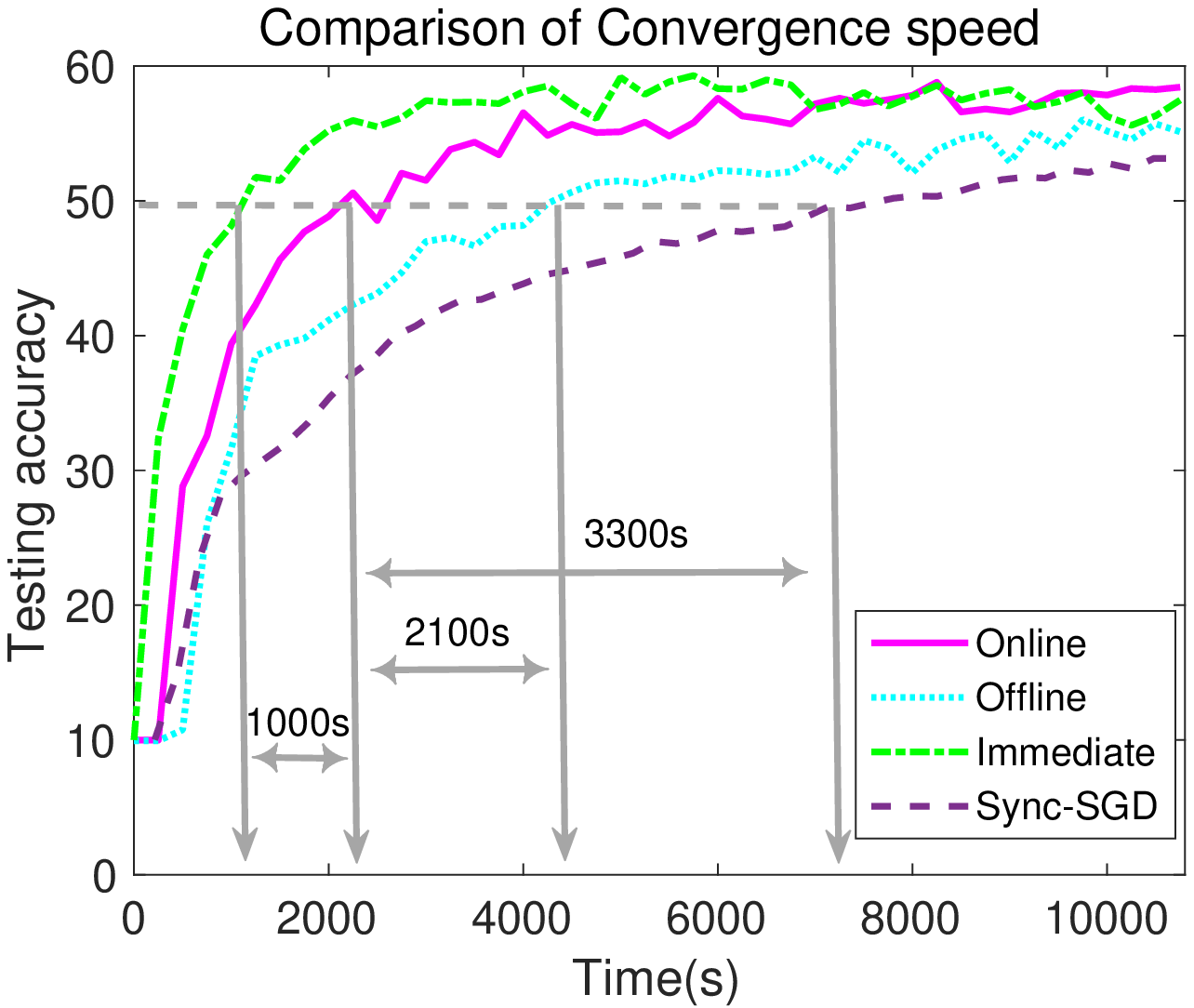}
                \vspace{-0.21in}
                \caption{}
\end{subfigure}
\begin{subfigure}[b]{0.245\textwidth}
                \includegraphics[width=1.02\textwidth]{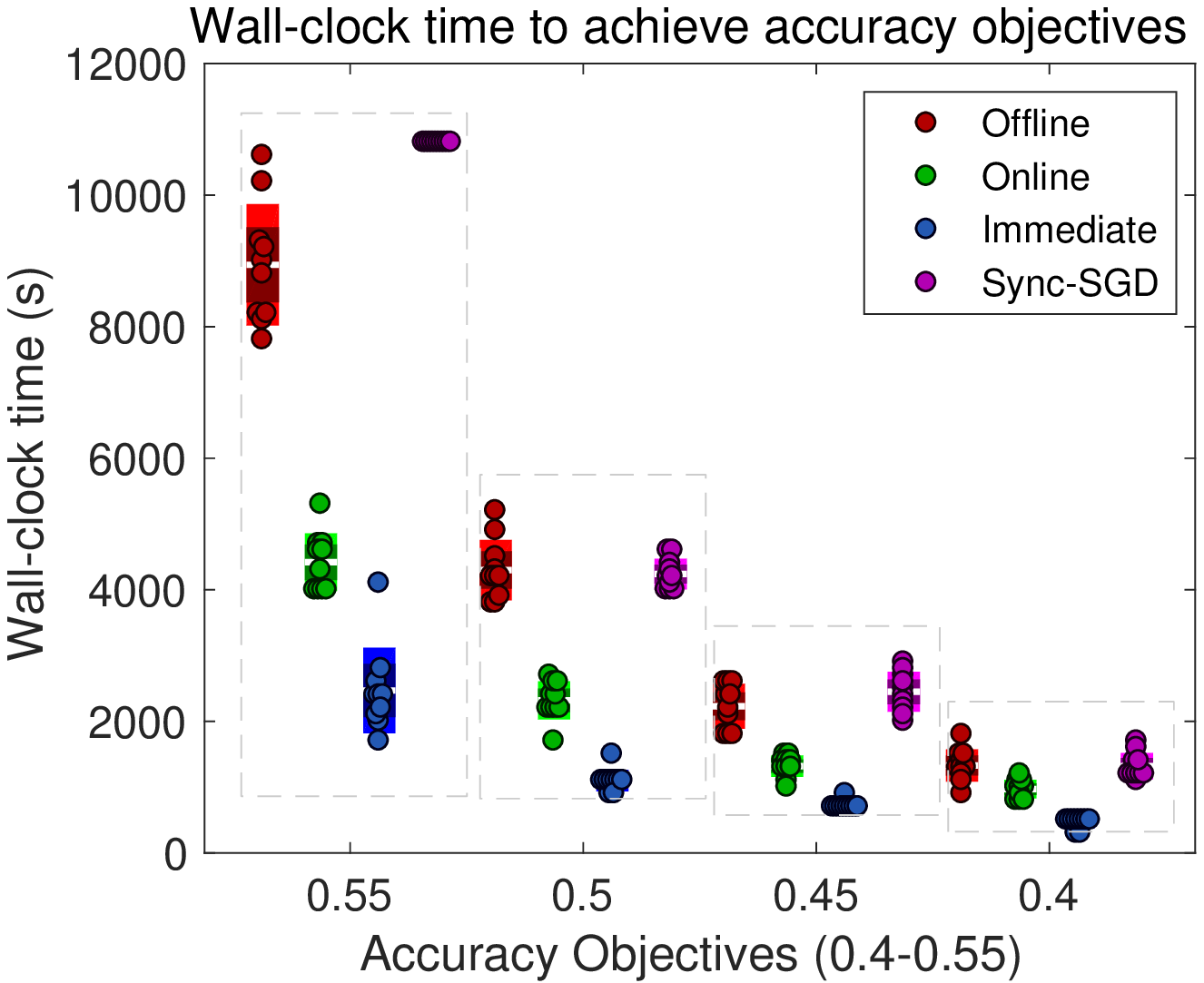}
                \vspace{-0.21in}
                \caption{}
\end{subfigure}
\begin{subfigure}[b]{0.245\textwidth}
                \includegraphics[width=1.02\textwidth]{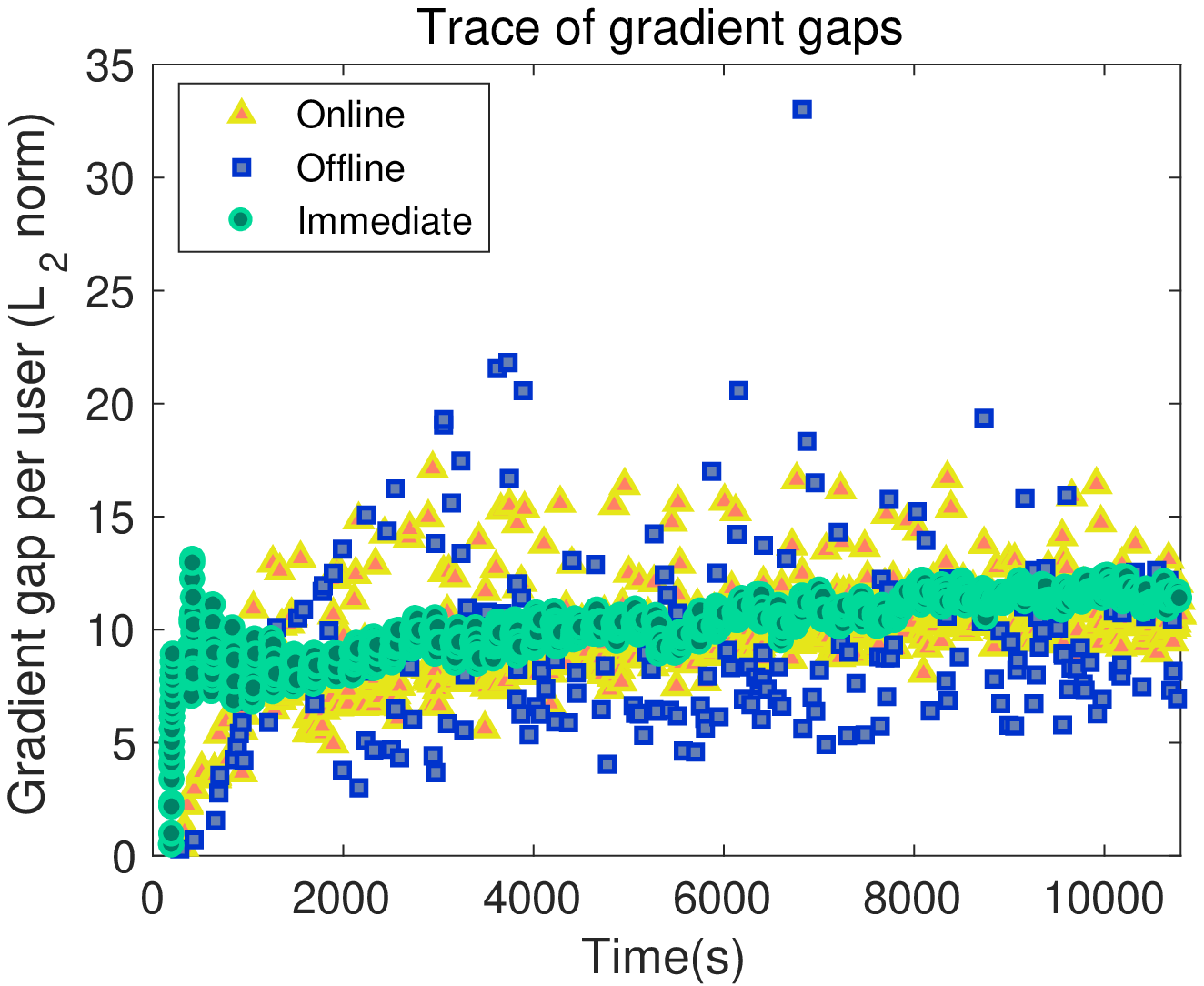}
                \vspace{-0.21in}
                \caption{}
\end{subfigure}%
\hspace*{-0.01in}
\caption{Comparison of convergence speed and gradient staleness (a) trace of gradient gap from Sync-SGD and ASync-SGD and the proportionality between lag and gradient staleness; (b) convergence speed of different schemes; (c) wall-clock convergence time to reach different accuracy objectives; (d) trace of gradient gaps from individual users.}
\label{convergence_fig}
\end{figure*}

\subsection{Energy Measurement}
First, we measure the energy consumptions of different control decisions as shown in Table \ref{energy_measurement}: \emph{training only} (1st row - considering training also as an app), \emph{application only} (1st col.) and \emph{co-running} (2nd col.). To mitigate the chances of breaking the devices while removing the battery and screen connectors, we use a combination of software profilers: Trepn~\cite{trepn} and Snapdragon Profiler~\cite{snapdragon} with the Monsoon Power Monitor~\cite{moonsoon}. Trepn is an old version for the last generation of Snapdragon chipsets (Nexus6/6P) and the newer version of Snapdragon Profiler supports the newer architectures of Pixel 2. For non-Snapdragon chipset (Hikey970), we directly power the development board with 12V DC input from the Monsoon Power Monitor.

We measure the system-wide energy consumption from the device which includes all the system background threads. To reduce the variances, we disable all irrelevant applications that might have processes lingering in the background. We choose a number of 8 popular applications that users usually spend considerable time. The percentage of energy saving is calculated as, $1 - \frac{P_i^{a'} t_a}{P_i^{b} t_b + P_i^{a} t_a}$ (notations from Eq. \eqref{power_eq}).

We notice that the newer generations of devices offer much higher energy saving ratio across all the applications (30-50\%) and a slightly increased execution time due to contentions of memory bandwidth. However, for the older chipset such as Nexus 6 with four homogeneous cores, co-running only offers marginal energy improvements depending on the application. Some applications even result an energy surge due to contention of cache resources, which further leads to CPU throttling and elongated training time. In these cases, the online controller is expected to avoid co-running.

\subsection{Simulation Evaluation}
\textbf{Evaluation Settings.} We set the probability of application arrival to $0.001$ in each time slot, i.e., an average of 1 app arrival for every 1000s. The application is chosen uniformly randomly from the 8 representative applications with the running time measured in Table \ref{energy_measurement}. We set the number of users to $25$ (equal partition of the CIFAR10 dataset) and each user randomly picks a device from the testbed. The total training time is set to 3 hours and each time slot is 1s. We compare the online algorithm against Sync-SGD~\cite{fedavg}, \emph{offline scheduling} (Knapsack Problem) and the fixed policy of \emph{immediate scheduling}, which runs the background training immediately when a device is available regardless of the application arrivals. We set the look-ahead time window of Knapsack evaluation to 500s with $L_b=1000$, which invokes the offline algorithm every 500s.

\textbf{Comparison of Energy Consumption.} Fig. \ref{energy_consumption}(a) compares the energy consumption of different scheduling policies. Immediate scheduling serves as an upper bound of energy consumption as it quickly turns on training regardless of the application arrival. In contrast, with the relaxed staleness bound $L_b=1000$, the Knapsack offline solution acts almost equivalently to a greedy scheme that is always waiting for co-running opportunities, thus incurs the minimum energy consumption. Reducing the value of $L_b$ would elevate the horizontal line of the offline solution. The online optimization evolves in the space in between: as $V$ grows, the energy consumption quickly drops and slowly approaches the offline solution around 200KJ, with a diminishing marginal gain when $V$ becomes large. The energy consumption stabilizes within an approximation factor of $1.14$ to the offline solution and 66\% energy savings compared to immediate scheduling and 63\% compared to Sync-SGD~\cite{fedavg}. Adjusting the staleness bound $L_b$ implies different levels of tolerance to staleness. With a larger $L_b$, more devices are put into the idling mode waiting for applications, thus the energy consumption is lower.

Figs. \ref{energy_consumption}(b)(c) show the queue backlogs $Q(t)$ and $H(t)$, which reflect the opposite of energy consumption in Fig. \ref{energy_consumption}(a). Both $Q(t)$ and $H(t)$ increase linearly after $V>10^4$ and this matches with Theorem 1 of Eq. \eqref{theorem1_2}. Fig. \ref{energy_consumption}(d) further validates the $[\mathcal{O}(1/V), \mathcal{O}(V)]$ energy-staleness trade-offs as the attempt of energy reduction in the systems would ultimately lead to congestion of the queues (high staleness and less update). An optimal choice of $V$ calls from the balance between energy and queue length backlogs -- $V$ around the value of $4000$, since further increasing $V$ beyond $V>10^4$ would have marginally reduced energy saving compared to the increase of queue length.

\begin{figure}[!t]
\centering
\hspace*{-0.2in}
\begin{subfigure}[b]{0.25\textwidth}
                \includegraphics[width=1.1\textwidth]{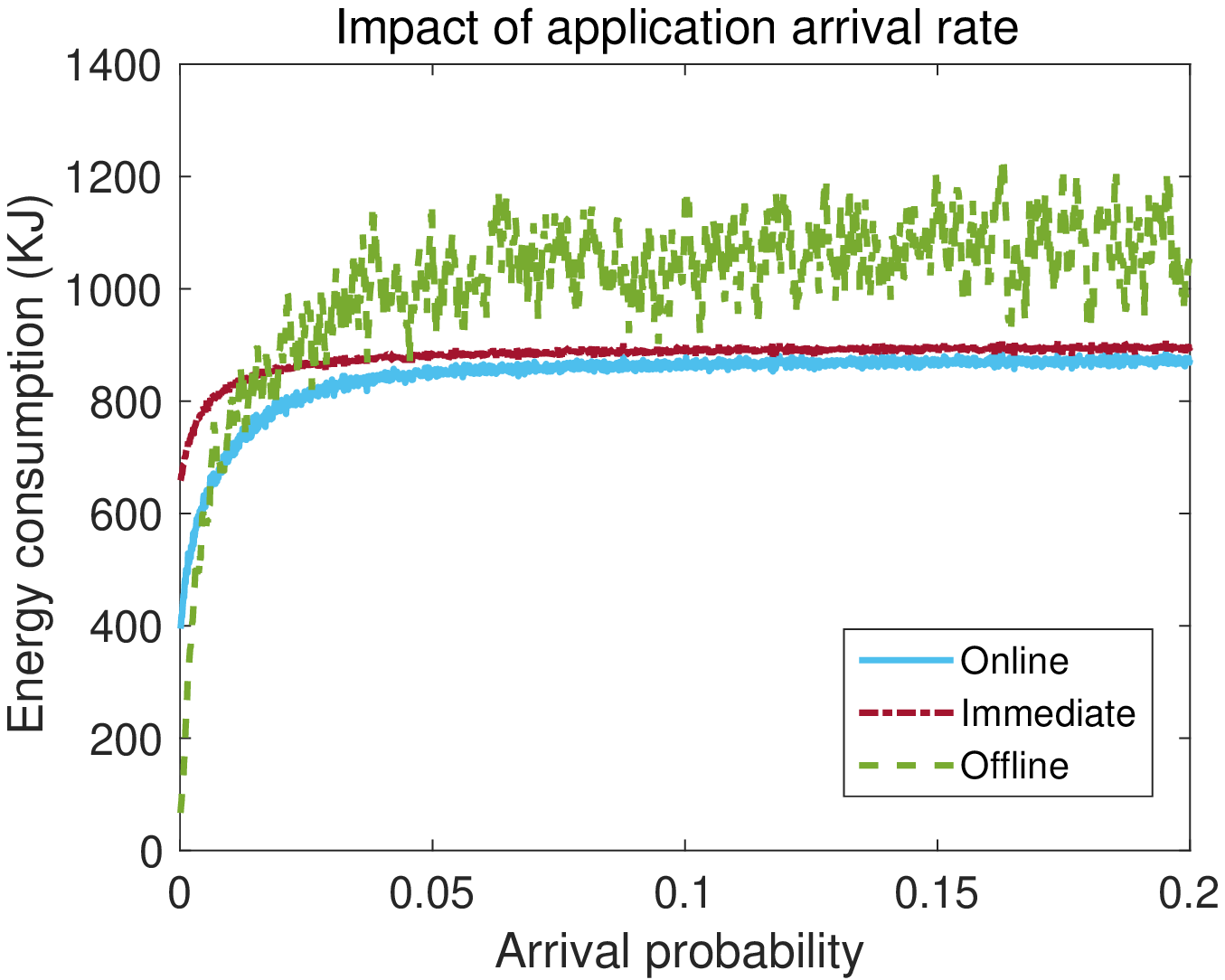}
                \vspace{-0.2in}
                \caption{}
\end{subfigure}
\begin{subfigure}[b]{0.25\textwidth}
                \includegraphics[width=1.1\textwidth]{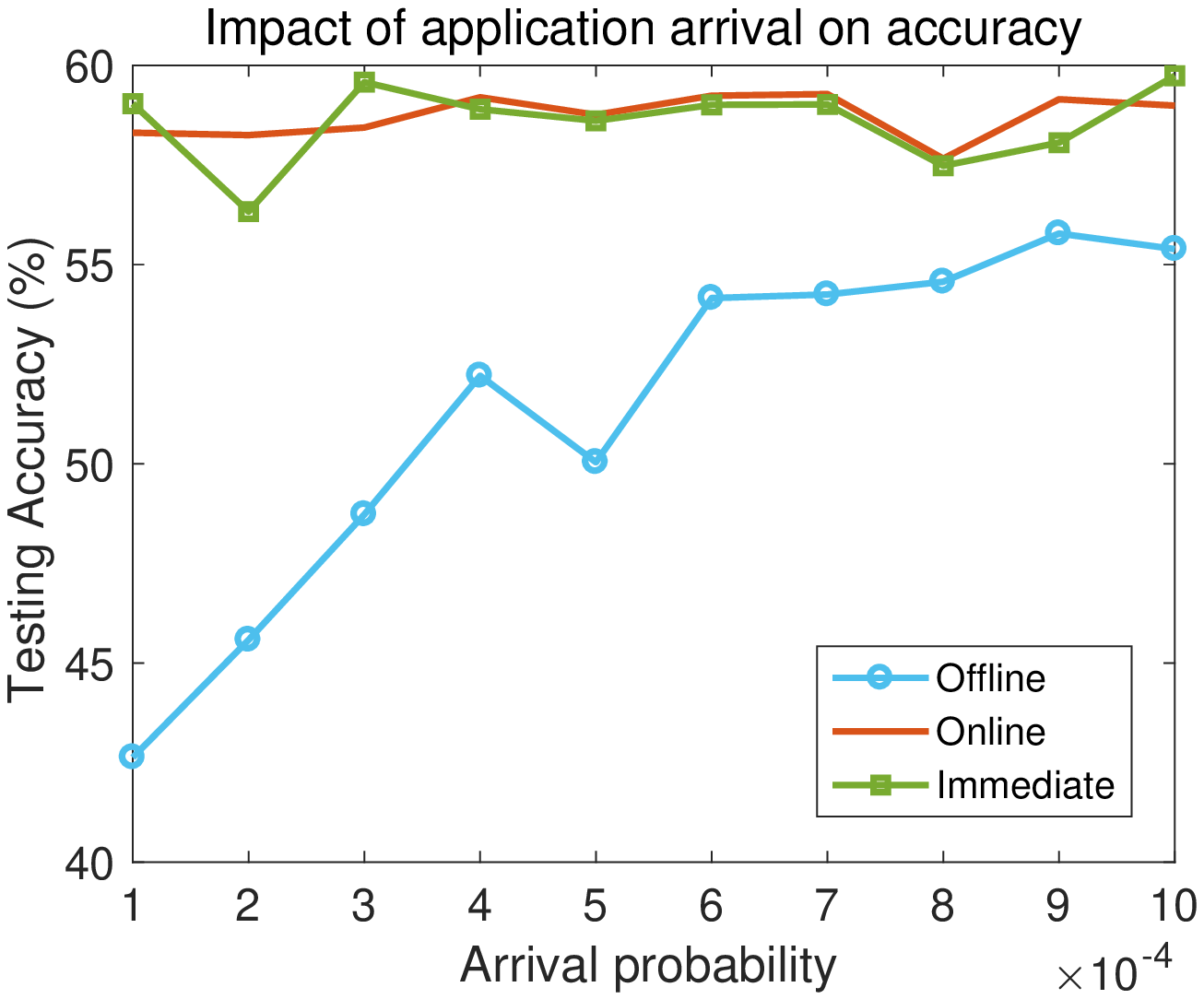}
                \vspace{-0.2in}
                \caption{}
\end{subfigure}
\caption{Impact of application arrival rate on (a) energy consumption; (b) testing accuracy with scarce application arrivals.}
\vspace{-0.1in}
\label{app_arrival}
\end{figure}

\textbf{Gradient Staleness and Convergence.} Since the control decisions further propagate upwards to affect the machine learning algorithms, the overall performance should be also measured in terms of how fast the model converges and the wall-clock time to reach a certain accuracy level. For ASync-SGD, the speed of convergence is dominated by: 1) the number of meaningful updates contributed by individual participants in fixed time intervals; 2) the accumulation and variance of gradient staleness. Fig. \ref{convergence_fig}(a) depicts the trace of gradient gaps of Sync-SGD and ASync-SGD when we fix the online algorithm with $V=4000$ and $L_b = 500$. The minimum values of Sync-SGD are sampled at the time of aggregation (model averaging), which follow a monotonically declining trend. Because of that, the gradient gaps from the local updates mostly stay in a narrow range and the variance is small. In contrast, the gradient gap from ASync-SGD forms an upward trend, especially at the beginning and the sideway movements in the later iterations lead to more fluctuations during convergence compared to Sync-SGD. The lower subplot also demonstrates a positive correlation between the simple count of updates (lag in Def. 1) and the gradient gap measured in norm difference between model parameters. Since the difference between local parameters tends to increase with more iterations, the same lag value could have different impact on staleness in different stages of training. To this end, gradient gap provides a more accurate measure for staleness.

Fig.\ref{convergence_fig}(b) shows the convergence of different approaches. If the objective for the model is to reach $0.5$ accuracy, the online scheme only lags the immediate scheduling by 1000s, but offers almost 60\% energy saving and both schemes ultimately converge to the same range of accuracy. The offline and Sync-SGD fall far behind mainly due to insufficient number of updates from the users. To compare the wall-lock training time more closely, we record the time for different schemes to reach 55\%, 50\%, 45\% and 40\% testing accuracy in Fig.\ref{convergence_fig}(c), by varying the random seeds to generate different random permutations of devices and application types. Since the testing accuracy of Sync-SGD has plateaued around 50\%, it never reaches 55\% during the 3-hour time frame. Similar to the previous results, Sync-SGD and offline scheme both result in the largest convergence time with different accuracy objectives. Immediate scheduling offers the fastest training with much higher cost of energy, especially when the accuracy objectives are lower (0.45, 0.4). The online scheme is capable of achieving a reasonable trade-off between energy consumption and training time.

Fig.\ref{convergence_fig}(d) shows the trace of gradient gaps for each individual user during training. As expected, the variance of the immediate scheduling is the smallest and the offline scheme results high variance in gradient staleness, which may lead to fluctuations of the testing accuracy. The variance of the online scheme evolves moderately in between as it does not either act too conservatively to withhold the users or too aggressively to activate them.

\textbf{Impact of Application Arrival.} Our strategy relies on the intensity of application usage for energy saving. We further evaluate the impact of different application arrival rates varying from $10^{-4}$ to $0.2$ per time slot, especially when the applications are scarce. Fig. \ref{app_arrival}(a) shows the application arrival rate vs. energy consumptions. With more running applications, the general energy consumption follows an increasing trend for all three schemes. Immediate scheduling is independent of application arrivals and the energy saving comes from the coincident co-running. In contrast, the online scheme is able to utilize the application arrival more wisely as we can see the initial gap from immediate scheduling is large. As the application rate rises, co-running quickly saturates and the online scheme has degraded into the immediate scheme. Because the offline scheme foresees all the co-running opportunities, it is able to achieve the lowest energy consumption when applications are scarce but will aggressively schedule with the applications when the arrival rate increases. Due to the random arrivals, the energy consumption has more variance with a larger arrival rate. As application usage depends on a variety of contextual cues such as time and location~\cite{app_usage}, it is highly possible that there is few application usage. Fig. \ref{app_arrival}(b) shows the impact on testing accuracy when applications are scarce. We can see that there is no noticeable accuracy degradation for the online scheme. Once the cost of the queue backlogs increases, the online scheme is able to switch back into the immediate scheme to clear the queue congestions. Thus, the offline scheme may offer better energy efficiency for different application rates, but the control decisions generate a negative feedback on the upper level convergence and testing accuracy when the applications are scarce. The online scheme provides more flexibility to adapt different application arrivals.

\textbf{Energy Overhead.} The online scheme examines Eq. \eqref{rhs_1}, which involves lightweight computation on the little cores. The energy overhead is shown in Table \ref{table_energy_overhead}, which is below 10\% in each time slot. To reduce the overhead, we can adjust the scheduling granularity. E.g., instead of making a decision in each time slot, we can enlarge the decision intervals, whereas this might miss co-running opportunities if the interval is larger than the application execution time. Due to space limit, we will demonstrate such trade-off in an extended version of this work.

\begin{table}[!ht]
\vspace*{-0.09in}
\centering
\small
\begin{tabular}[t]{l c c c c}
\hline
&Nexus 6&Nexus 6P &Pixel 2\\
\hline
Power(idle) &0.238  &0.486  &0.689   \\
Power(comp.)  &0.245  &0.525  &0.736   \\
Overhead (\%) &3.0\%  &7.4\%   &6.3\%        \\
\hline
\end{tabular}
\caption{Energy overhead of online optimization (W).} \label{table_energy_overhead}
\end{table}%

\vspace{-0.05in}
\section{Conclusion and Future Work} \label{sec:conclusion}
In this paper, we combine ASync-SGD and application co-running to minimize energy consumptions of federated tasks on mobile devices. We motivate this work by real measurements and illustrate the fundamentals of energy saving. Then we develop the offline and online schemes to explore the energy-staleness trade-off with low computational overhead. Our extensive evaluation demonstrates that the online optimization achieves over 60\% energy saving compared to the benchmarks, and only 15\% away from the offline solution.

The proposed mechanism can adapt to different diurnal and nocturnal application usage patterns by taking advantage of the common temporal activities from the users, while keeping the devices in low power state during the rest of the time. Though we only demonstrate the convergence empirically, in principle, the theoretical convergence is guaranteed given a bounded delay of gradients~\cite{ijcai16}. The Lyapunov framework manages the delay from the virtual queue bounded by the gradient staleness $L_b$ in Eq. \eqref{constraint21}. We defer the rigorous theoretical proofs to the future works.

\section{Acknowledgement}
We are grateful to the anonymous reviewers for their insights and detailed comments, and the support from the U.S. National Science Foundation under grant number 2152580 and 2007386. 

\section{Appendix}   \label{appendix}

\subsection{Proofs of Lemma 2}  \label{appendix-1}
Applying Eq. \eqref{queueing_all} to Eq. \eqref{drift_eq}, we have,
\begin{eqnarray}
\small
\Theta(t) &=& \mathbb{E}\{ L(\Theta(t+1)) - L(\Theta(t))|\Theta(t)\} \nonumber \\
&=& \frac{1}{2} \mathbb{E}\{ \Theta(t+1)^2 - \Theta(t)^2 \}  \nonumber\\
&=& \frac{1}{2} \mathbb{E}\{ Q(t+1)^2 - Q(t)^2 + H(t+1)^2 - H(t)^2 \}  \notag \\ & \label{proof1_1}
\end{eqnarray}
Since $\max^2\{x,0\} \leq x^2$, from Eq. \eqref{queueing_eq1} and Eq. \eqref{queueing_eq2} we have,
\begin{eqnarray}
\small
&&Q^2(t+1) + H^2(t+1) \leq Q^2(t) + (A(t) - b(t))^2 + \nonumber\\
&&2 Q(t)(A(t) - b(t)) + H^2(t) + G(t)^2 + 2 H(t)G(t) + L_b^2  \notag \\ & \label{proo1_2}
\end{eqnarray}
From \eqref{proof1_1},
\begin{equation}
\small
\Theta(t) \leq  B + \mathbb{E} \{Q(t) (A(t) - b(t)) | Q(t)\} + \mathbb{E} \{H(t)G(t)| H(t)\}  \label{proof1_3}
\end{equation}
where $B = \frac{1}{2} (A_{max}^2 + B_{max}^2 + G^2_{max} + L_b^2)$. $A_{max}, B_{max}$ and $G_{max}$ are the maximum arrival, service rate and gradient gap in the system. Thus, Eq. \eqref{proof1_3} completes the proof of \emph{Lemma 2}.

\subsection{Proofs of Theorem 1} \label{appendix-2}
For the optimal decision $\alpha^\ast(t)$ that can stabilize the queue,
\begin{equation}
\small
\mathbb{E}\{P(\alpha^\ast(t))\} = P^\ast.  \label{optimality}
\end{equation}
Since we can adjust the control decisions, there must exist $\epsilon_1, \epsilon_2 > 0$ so the difference between the service and arrival rates of the actual and virtual queues are larger than $\epsilon_1, \epsilon_2$ respectively (considering $L_b$ as the fixed service rate of the virtual queue $H(t)$):
\begin{eqnarray}
\small
&&\mathbb{E}\{b_i(t) - A_i(t)|Q(t)\} > \epsilon_1 \label{optimality_epsilon1} \\
&&\mathbb{E}\{L_b - g_i(t,t+\tau)|H(t)\} > \epsilon_2 \label{optimality_epsilon2}
\end{eqnarray}
Plugging Eqs. \eqref{optimality_epsilon1} and \eqref{optimality_epsilon2} into Eq. \eqref{lemma2},
\begin{equation}
\small
\Delta (\Theta(t)) + V \mathbb{E}\{P(t)| \Theta(t)\}  \leq B + V P^\ast - \epsilon_1 \mathbb{E}\{Q(t)\} - \epsilon_2 \mathbb{E}\{H(t)\} \label{lemma1_cp}
\end{equation}
Taking the summation over $t \in \{0, \cdots, T-1\}$,
\begin{eqnarray}
\small
&&\sum\limits_{t=0}^{T-1}\Delta (\Theta(t)) + \sum\limits_{t=0}^{T-1} V \mathbb{E}\{P(t)| \Theta(t)\}  \leq T( B + V P^\ast) \nonumber \\
&& - \sum\limits_{t=0}^{T-1} (\epsilon_1 \mathbb{E}\{Q(t)\} + \epsilon_2 \mathbb{E}\{H(t)\}) \label{lemma1_cp2}
\end{eqnarray}
Plugging Eq. \eqref{drift_eq} into Eq. \eqref{lemma1_cp2}, and dividing both sides by $T \cdot V$,
\begin{eqnarray}
\small
&&\frac{\mathbb{E}\{ L(\Theta(T-1)) - L(\Theta(0))\}}{T V} + \frac{1}{T} \sum\limits_{t=0}^{T-1} \mathbb{E}\{P(t)\}  \leq \frac{B}{V} \nonumber \\
&& + P^\ast - \frac{\epsilon_1}{T V} \mathbb{E}\{Q(t)\} - \frac{\epsilon_2}{T V} \mathbb{E}\{H(t)\} \label{lemma1_cp3}
\end{eqnarray}
Since $L(\Theta(0)) = 0$, when $T \rightarrow \infty$, we can simplify Eq. \eqref{lemma1_cp3} as,
\begin{equation}
\small
\limsup\limits_{T \rightarrow \infty} \frac{1}{T} \sum\limits_{t=0}^{T-1} \mathbb{E}\{P(t)\}  \leq  \frac{B}{V} + P^\ast \label{lemma1_cp4}
\end{equation}
The time averaged queue length can be derived by dividing $\epsilon T$,
\begin{eqnarray}
\small
&&\frac{1}{T} \sum\limits_{t=0}^{T-1} \mathbb{E}\{\Theta(t)\} \leq \frac{B+ V (P^\ast - \frac{1}{T}\sum\limits_{t=0}^{T-1} \mathbb{E}\{P(t)\}) }{\epsilon} \nonumber \\
&& - \frac{(\sum\limits_{t=0}^{T-1} \epsilon_1 \mathbb{E}\{Q(t)\} + \epsilon_2 \mathbb{E}\{H(t)\})}{\epsilon T} + \frac{\mathbb{E}\{L(\Theta(0))\}}{\epsilon T} \nonumber \\ \label{lemma1_cp5}
\end{eqnarray}
Taking the limits of $T \rightarrow \infty$,
\begin{equation}
\small
\overline{\mathbf{\Theta}} = \limsup\limits_{T \rightarrow \infty} \frac{1}{T} \sum\limits_{t=0}^{T-1} \mathbb{E}\{\Theta(t)\}  \leq  \frac{B}{\epsilon} + \frac{V (P^\ast - \overline{P})}{\epsilon} \label{lemma1_cp6}
\end{equation}


\end{document}